\documentclass[useAMS,usenatbib]{mn2e}

\usepackage{graphicx}
\usepackage{amsmath}
\usepackage[dvips,usenames]{color}

\def\aj{AJ}%
\def\araa{ARA\&A}%
\def\apj{ApJ}%
\def\aap{A\&A}%
\def\aaps{A\&AS}%
\def\mnras{MNRAS}%
\def\pasp{PASP}%
\newcommand{\realfigure}[3]{
              \begin{figure}
              \includegraphics[width=84mm]{#1}
              \caption{#2}
              \label{#3}
              \end{figure}}

\newcommand{\realfiguretwofig}[4]{
              \begin{figure*}
              \centering
	          \includegraphics[width=84mm]{#1}\hfill
	          \includegraphics[width=84mm]{#2}
              \caption{#3}\label{#4}
              \end{figure*}}
              
\newcommand{\realfigurevert}[4]{
              \begin{figure}
              \centering
	          \includegraphics[width=84mm]{#1}\hfill
	          \includegraphics[width=84mm]{#2}
              \caption{#3}\label{#4}
              \end{figure}}

\newcommand{\mytab}[6]{
\begin{table}
\caption{#4}
\label{#5}
\begin{center}
\begin{#6}
\begin{tabular}{#1}
\hline 
#2
\hline
#3
\hline
\end{tabular}
\end{#6}
\end{center}
\end{table}
}

\newcommand{\mytabbig}[6]{
\begin{table*}
\begin{minipage}{126mm} 
\caption{#4}
\label{#5}
\begin{center}
\begin{#6}
\begin{tabular}{#1}
\hline 
#2
\hline
#3
\hline
\end{tabular}
\end{#6}
\end{center}
\end{minipage}
\end{table*}
}

\newcommand{\leoii}{\mbox{Leo\,II}}

\newcommand{\feh}{{\rm [Fe/H]}}
 \newcommand{\cah}{{\rm [Ca/H]}}
\newcommand{\mh}{{\rm [M/H]}}

\newcommand{\vks}{\mbox{$V\!-\!K_s$}}
\newcommand{\jks}{\mbox{$J\!-\!K_s$}}
\newcommand{\jhs}{\mbox{$J\!-\!H_s$}}
\newcommand{\bv}{\mbox{$B\!-\!V$}}
\newcommand{\bks}{\mbox{$B\!-\!K_s$}}
\newcommand{\ebv}{\mbox{$E_{B\!-\!V}$}}
\newcommand{\dmo}{\mbox{$(m\!-\!M)_{0}$}}
\newcommand{\Teff}{\mbox{$T_{\rm eff}$}}
\newcommand{\Msun}{\mbox{$M_{\odot}$}}

\newcommand{\caii}{Ca\,{\sc ii}}

\newcommand{\referee}[1]{}         

\newcommand{\abbrev}[1]{{#1}}      

\newcommand{\sfhrizzi}{Rizzi et al. (in prep.)}
\newcommand{\sfhrizzip}{Rizzi et al., in prep.}

\title[
The evolved stars of \leoii\ dSph in the near-IR
]{
The evolved stars of \leoii\ dSph galaxy from near-infrared 
UKIRT/WFCAM observations }

\author[Gullieuszik et al.]{
M. Gullieuszik$^{1}$\thanks{E-mail:
marco.gullieuszik@oapd.inaf.it},
E.~V.~Held$^{1}$,
L.~Rizzi$^{2}$,
L.~Girardi$^{1}$,
P.~Marigo$^{3}$ and
Y.~Momany$^{1}$
\\
$^{1}$INAF/Osservatorio Astronomico di Padova,
vicolo dell'Osservatorio 5, I-35122 Padova, Italy
\\
$^{2}$Joint Astronomy Centre, 660 N. A'ohoku Place, 
University Park, Hilo, HI 96720, USA
\\
$^{3}$Dipartimento di Astronomia, Universit\`a di Padova, 
vicolo dell'Osservatorio 2, I-35122 Padova, Italy
}

\begin{document}

\date{Accepted \dots  Received \dots ; in original form \dots }

\pagerange{\pageref{firstpage}--\pageref{lastpage}} \pubyear{2008}

\maketitle

\label{firstpage}

\begin{abstract}
We present a study of the evolved stellar populations in the dwarf
spheroidal galaxy \leoii, based on $JHK_s$ observations obtained with
the near-infrared array WFCAM at the UKIRT telescope. 
Combining the new data with optical data, we derived photometric
estimates of the distribution of global metallicity \mh\ of individual
red giant stars from their \vks\ colours. Our results are consistent
with the metallicities of RGB stars obtained from \caii\ triplet
spectroscopy, once the age effects are considered.  The photometric
metallicity distribution function has a peak at \mh~$=-1.74$
(uncorrected) or \mh~$=-1.64 \pm 0.06$ (random) $ \pm 0.17$ (systematic)
after correction for the mean age of \leoii\ stars (9 Gyr). The
distribution is similar to a Gaussian with $\sigma_{\rm [M/H]}=$~0.19
dex, corrected for instrumental errors.  
We used the new data to derive the properties of a nearly complete
sample of asymptotic giant branch (AGB) stars in \leoii.  Using a
near-infrared two-colour diagram, we were able to obtain a clean
separation from Milky Way foreground stars and discriminate between
carbon- and oxygen-rich AGB stars, which allowed \referee{us} to study their
distribution in $K_s$-band luminosity and colour.
We simulate the $JHK_{\rm s}$ data with the {\sc trilegal} population
synthesis code together with the most updated thermally pulsing AGB
models, and using the star formation histories derived from independent
work based on deep HST photometry.
After scaling the mass of \leoii\ models to the observed number of upper
RGB stars, we find that present models predict too many O-rich TP-AGB
stars of higher luminosity due to a likely under-estimation of either
their mass-loss rates at low metallicity, and/or their degree of
obscuration by circumstellar dust.
On the other hand, the TP-AGB models are able to reproduce the observed
number and luminosities of carbon stars satisfactorily well, indicating
that in this galaxy the least massive stars that became carbon stars
should have masses as low as $\sim 1\, M_{\odot}$.
\end{abstract}

\begin{keywords}
Galaxies: individual: \leoii\  -- 
Galaxies: stellar content --
stars: AGB and post-AGB -- 
stars: carbon -- 
Local Group
\end{keywords}

\section{Introduction}

\leoii\ is one of the most distant dwarf spheroidal (\abbrev{dSph})
satellites of the Milky Way.  A number of photometric studies derived
quite different distances for \leoii.  \citet{mighrich1996},
using the $V$  magnitude of the horizontal branch (\abbrev{HB}),
placed the galaxy 
at a distance modulus $(m-M)_0=21.55\pm0.18$.  Using the $I$
band magnitude of the tip of the RGB (TRGB), \citet*{bella+2005} found
$(m-M)_0=21.84\pm0.13$.
Distance estimates in the literature are intermediate between these
values, with typical uncertainties of $\sim0.2$ magnitudes. 

The estimates of mean metallicity of \leoii\ range from \feh\ $=-1.6$
\citep{mighrich1996} up to \feh\ $\simeq-1.1$ \citep{dolp2002}.
Recently, two independent spectroscopic studies, based on the \caii\
triplet method, found \leoii\ to be relatively metal-poor: a mean value
of \feh\ $=-1.74$ was derived by \citet{koch+2006leo2} while
\citet*{bosl+2007} found \feh~$=-1.59$. 

The \leoii\ dSph was considered for a long time as a typical ``old''
dSph.  Then \citet{mighrich1996} obtained an HST/WFPC2 colour magnitude
diagram (CMD) reaching about 2 magnitudes below the oldest main sequence
turnoff.  By analysing the distribution of stars near the base of the
red giant branch (RGB), they determined that the first generation of
stars in \leoii\ was coeval with the formation of Galactic Globular
clusters, and nearly half of the stars formed during a period of star
formation lasting about 4 Gyr, with the typical star forming 
about 9 Gyr ago.
They found a negligible rate of star formation in the last $\sim7$
Gyr. Subsequent re-analyses of these data confirmed this scenario and
found a star formation history (\abbrev{SFH}) dominated by old stellar
populations with a low star formation rate in the last 8 Gyr
\citep{hern+2000,dolp2002,dolp+2005}.
A recent wide-field study has shown a gradient in the HB morphology and
the mean age of stellar populations, with a significant population
younger than 8 Gyr found only at the centre \citep{komi+2007}.

The presence of a small intermediate age population in \leoii\ is also
indicated by a small number of C stars
\citep*{aaro+1983,aaro+moul1985,azzo+1985}. \citet{azzo+1985} list 
six certain C stars and one candidate, while \citet{azzo2000} stated
they found two new ones but without providing further details. 

As part of an imaging study of Local Group (LG) dwarf galaxies in the
near infrared (\abbrev{NIR}), we have undertaken a study of the evolved
stellar populations in \leoii\ using wide-area NIR imaging. The main
goals are to study the metallicity distribution of red giant stars and
to obtain $J$, $H$, and $K_s$ magnitudes of asymptotic giant branch
(AGB) stars \citep{gull+2007for,gull+2007sag}.
The main advantage of NIR observations for studying AGB stars over the
alternative search technique based on intermediate band imaging in the
optical \citep*{albe+2000,battdeme2000,nowo+2001}, is that the spectral
energy distribution of cool AGB stars peaks in the NIR
\citep[e.g.,][]{gull+2007for}.
Also, bolometric corrections are smaller and more precise in the NIR
making comparison with theoretical quantities easier.  In addition, the
(foreground and internal) extinction is much lower in the NIR than in
the optical \citep{rieklebo1985}.

\section{Observations and reduction}\label{s:leo2obs}

\mytab{
cccccc}{
Filt.&
N$_{\text{ima}}$&
DIT(s)&
N$_{\text{exp}}$&
N$_{\text{jit}}$&
microsteps\\}{
$J$  & 6 & 5.0 & 2 & 9 & $2\times2$ \\
$H$  & 6 & 5.0 & 2 & 9 & $2\times2$\\
$K_s$ & 10 & 5.0 & 2 & 9 & $2\times2$\\}{
Observing Log}{
t:leo2obslog}{
normalsize}

Observations of \leoii\ were carried out in April 19--20, 2005 using the
wide-field NIR camera WFCAM at the UKIRT telescope on Mauna Kea,
Hawaii.  WFCAM uses 4 Rockwell Hawaii-II HgCdTe detectors with a pixel
scale of $0\farcs4$ pixel$^{-1}$. The size of each array is
$2048\times2048$ pixels, corresponding to about $13\farcm6 \times
13\farcm6$, and the 4 arrays are separated by a gap comprising 
94\% of the detector size.
Since the area encompassed by one array is comparable to the galaxy size
\referee{\citep[the tidal radius of \leoii\ is 
$\sim 9\arcmin$,][]{cole+2007,komi+2007}},
we used only one array (No.~3) to
observe \leoii.  The remaining detectors provide a useful estimate of
the foreground/background counts.

Each ``observation'' (or pointing) consisted of 4 micro-stepped 10\,s
images (two 5\,s coadds) on a 9-points jitter pattern, totalling a 360\,s
exposure time.  In total, 6 observations were obtained in $J$ and $H$
and 10 in the $K_s$ band, yielding a total on-target integration time of
36\,min in $J$ and $H$ and 1\,h in $K_s$.  The $2\times2$ ``small''
microstepping was used to obtain a better spatial sampling.  
The observing log is given in Table~\ref{t:leo2obslog}.

The raw data were processed using the WFCAM pipeline provided by the
VISTA Data Flow System Project, to which the reader is referred for
details \citep{dye+2006}. The pipeline combines the micro-stepped images
in each band into 4k$\times$4k ``Leavstack'' oversampled images with a
spatial resolution twice that of the original raw images,
i.e. $0\farcs2$ pixel$^{-1}$.  The pipeline products are astrometrically
calibrated using the ZPN projection \citep{cala+grei2002} and the 2MASS
Point Source Catalogue \citep[PSC,][]{skru+2006} as a reference, with a
final systematic accuracy of the order $0\farcs1$.
In our analysis, we made use only of the array where \leoii\ had been
centred (No.~3) along with a second array (No.~2) used to estimate the
contribution of the foreground Galactic stars and background galaxies.

Point spread function (\abbrev{PSF}) photometry was performed on the
individual oversampled images (6 in $J$ and $H$, 10 in $K_s$) using the
{\sc allstar}/{\sc allframe} \citep{stet1987,stet1994}.  The PSF was
generated with a Penny function with quadratic dependence on the
position on the frame. The final catalogue includes instrumental PSF
magnitudes of objects detected in at least 2 images in 2 bands.

The positions of the sources in the raw photometric catalogue were
converted from pixels to the J2000 equatorial system using the
astrometric calibration provided by the pipeline and IRAF\footnote{The
Image Reduction and Analysis Facility (IRAF) software is provided by the
National Optical Astronomy Observatories (NOAO), which is operated by
the Association of Universities for Research in Astronomy (AURA), Inc.,
under contract to the National Science Foundation.} tasks including
support for the ZPN projection.

\realfigure{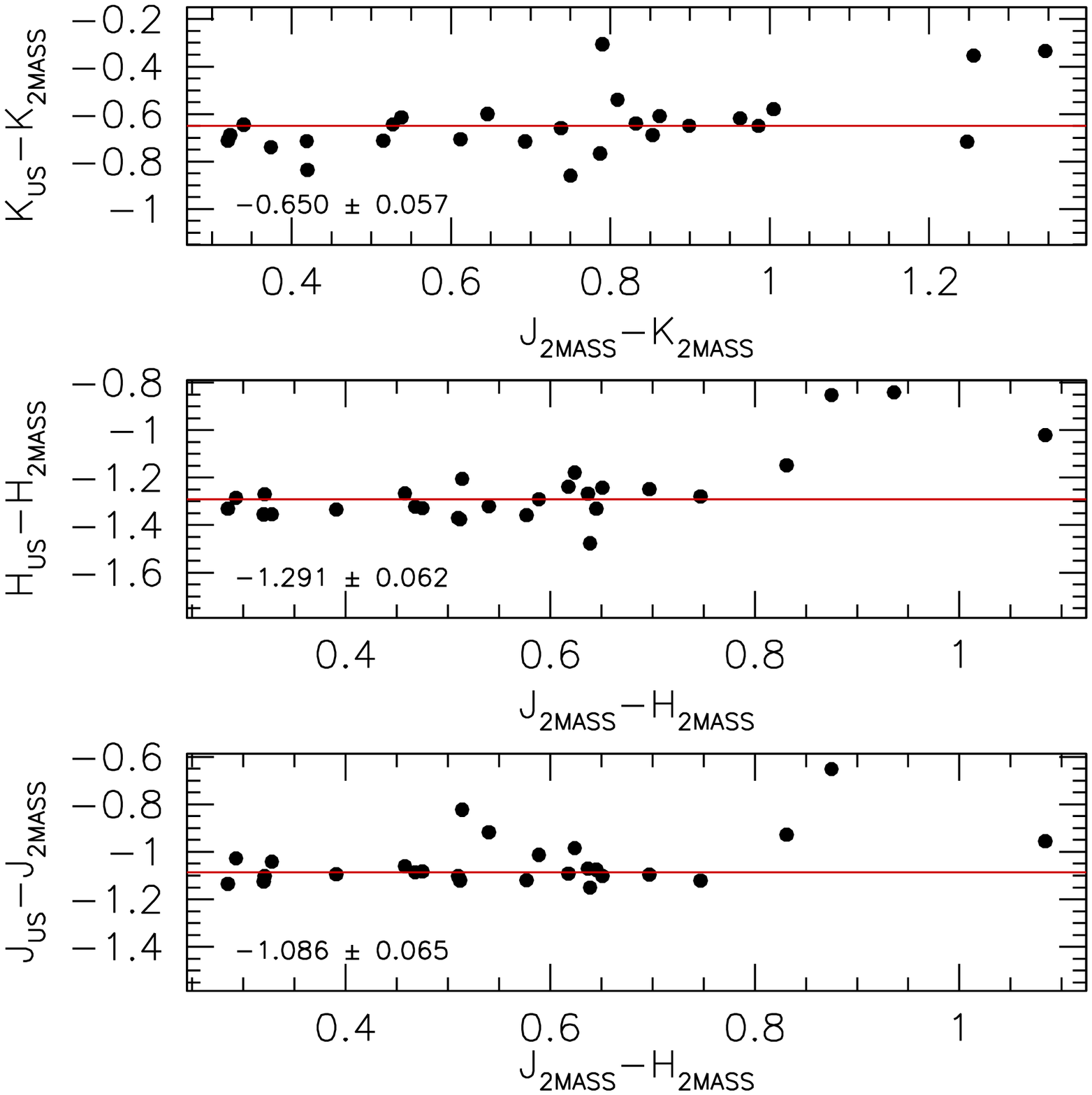}{Comparison of \leoii\ WFCAM
  instrumental magnitudes corrected using the \citet{dye+2006} colour
  terms with 2MASS photometry, for stars in common with the 2MASS PSC.
  No residual colour terms are detectable.  In each panel the median
  difference and the standard deviation of the data are
  shown.}{f:leo2calib2mass}

\referee{ 
  Our raw photometric catalogue was calibrated onto the system defined
  by the 2MASS PSC, by applying the colour terms between the WFCAM and
  2MASS systems derived by \citet{dye+2006}.
  The $JHK_s$ photometric zero points were then calculated by
  comparing our transformed magnitudes with the 2MASS magnitudes for 
  stars in common with the PSC.  Only 2MASS stars with S/N ratio
  greater than 10 and a photometric error smaller than 0.1 mag were
  used.
  The magnitude differences in $JHK_s$ are plotted in
  Fig.~\ref{f:leo2calib2mass}.  We calculated the median zero point
  shifts and standard deviations using a $\kappa$-$\sigma$ clipping to
  exclude the outliers. The r.m.s. error of the residuals is $\sim
  0.06$ mag in all bands, a value that we take as the uncertainty on
  our absolute photometric calibration. The median shifts were used to
  tie our photometry to the 2MASS system.
  The large scatter of the reddest stars, with $\jks>1.0$ and
  $\jhs>0.8$, can be explained by the fact that they are AGB stars in
  \leoii\ (as shown in the next Section), which are mostly long-period
  variable stars \citep[see, e.g.,][]{whit+2006}, and are
  close to the magnitude limit of the 2MASS.
}
The final, calibrated NIR photometric catalogue of stars in
\leoii\ is provided in the electronic version of the journal.  A few
lines are presented in Table~\ref{t:leo2catal} to illustrate its
content.

\realfigure{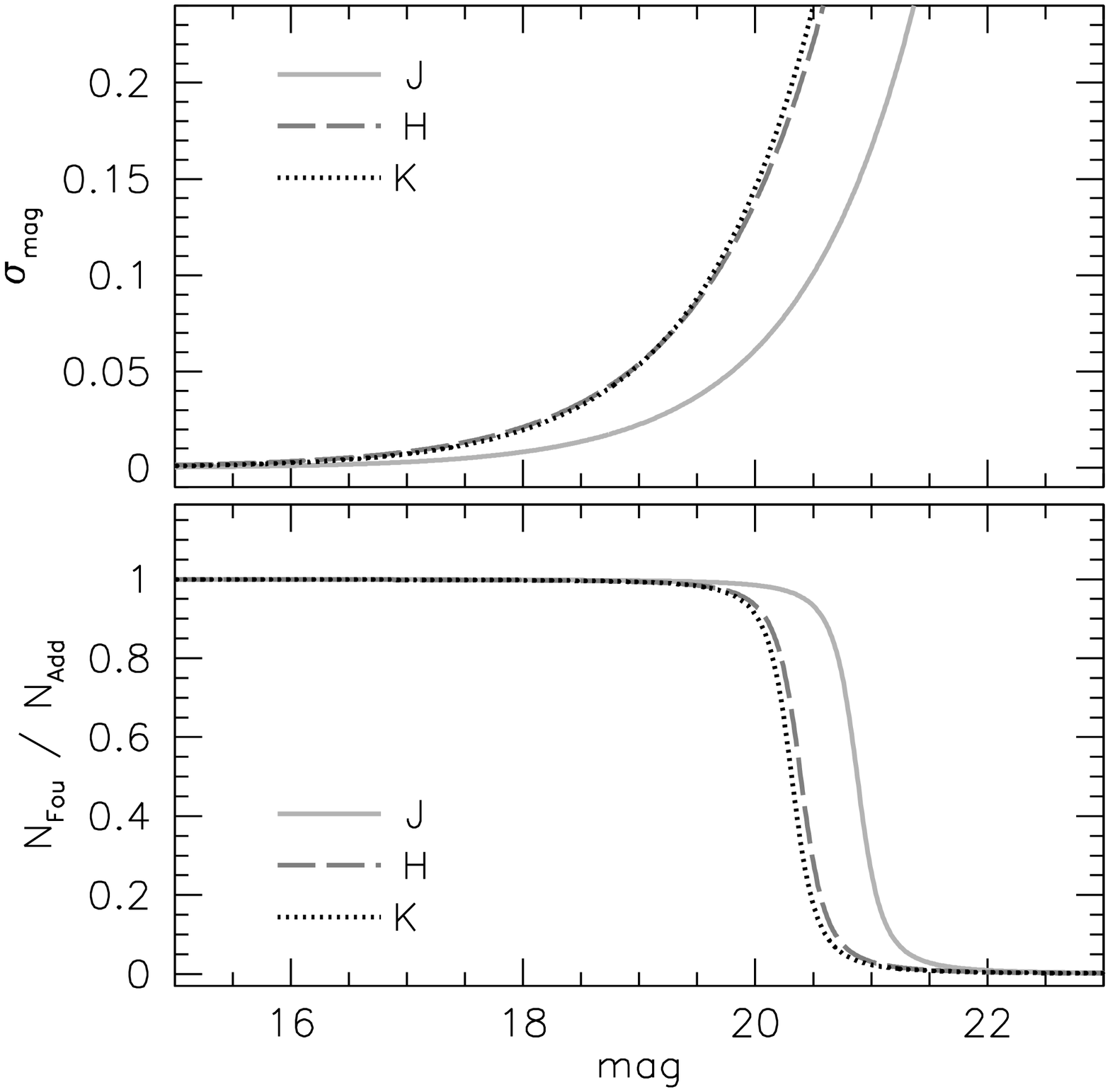}{Completeness ({\it lower panel}) and
  photometric errors ({\it upper panel}) in our \leoii\ images as
  derived from artificial star experiments, plotted against the input
  magnitudes of simulated stars.}{f:compl}

Artificial star experiments were also performed to evaluate the
photometric errors and completeness of our photometry.  
We performed 20 test runs adding $\sim2000$ stars on a $2k \times 2k$
portion of the frames.  The input magnitudes and colours were randomly
generated to reproduce the RGB of \leoii.  The results of our
experiments (completeness factor and internal r.m.s. photometric errors)
are shown in Fig.~\ref{f:compl}.
The completeness factor is larger than 50\% for magnitudes brighter than
$K_s \simeq 20$.  Note, however, that most results of this paper are
based on photometry of stars brighter than $K_s \simeq 18$, for which we
have a completeness factor $\simeq 100\%$ and photometric errors smaller
than 0.02 mag.

Possible spatial variations in the completeness factor and photometric
errors were investigated by repeating the analysis of artificial star
experiments for different regions of our frames, and no significant
variation was found.  In fact, our scientific photometric catalogue only
contains $\sim 3200$ objects, hence we expect spatially-varying
crowding effects to be unimportant.
The effect of a photometric bias towards brighter retrieved magnitudes,
caused by photometric blends \citep[e.g.,][]{gall+1996}, was also
investigated and found negligible (in our experiments the mean
difference is always less than 0.01 mag for stars brighter than $K_s
\simeq 18$).

\begin{table}
\caption{
The NIR catalogue of \leoii\ stars over WFCAM array No.~3. A
few lines are shown here for guidance regarding its form and content,
while the full catalogue is available from the electronic edition of the
journal. \referee{See Fig.~\ref{f:compl} for photometric errors}.}
\label{t:leo2catal}
\begin{center}
\begin{tabular}
{l @{\hspace{3.5mm}}c @{\hspace{3.5mm}}c @{\hspace{3.5mm}}c @{\hspace{3.5mm}}c @{\hspace{3.5mm}}c }
\hline
ID & $\alpha$ (J2000) & $\delta$ (J2000) & $J$ & $H$ & $K_s$ \\
\hline                    
 1 & 11:13:26.53 & +22:02:09.7 & 21.01 & 19.38 & 19.41\\
 2 & 11:13:11.40 & +22:02:10.2 & 19.97 & 19.13 & 18.26\\
 3 & 11:13:11.68 & +22:02:12.1 & 20.00 & 19.13 & 18.25\\
 4 & 11:13:45.11 & +22:02:14.3 & 20.55 & 19.21 & 18.45\\
 5 & 11:13:29.49 & +22:02:19.0 & 20.63 & 19.81 & 18.75\\
\hline                  
\end{tabular}
\end{center}
\end{table}

\section{Colour-Magnitude diagrams}\label{s:leo2cmd}

\begin{figure*}
\begin{center}
\begin{tabular}{c c}
\includegraphics[clip,width=.45\textwidth]{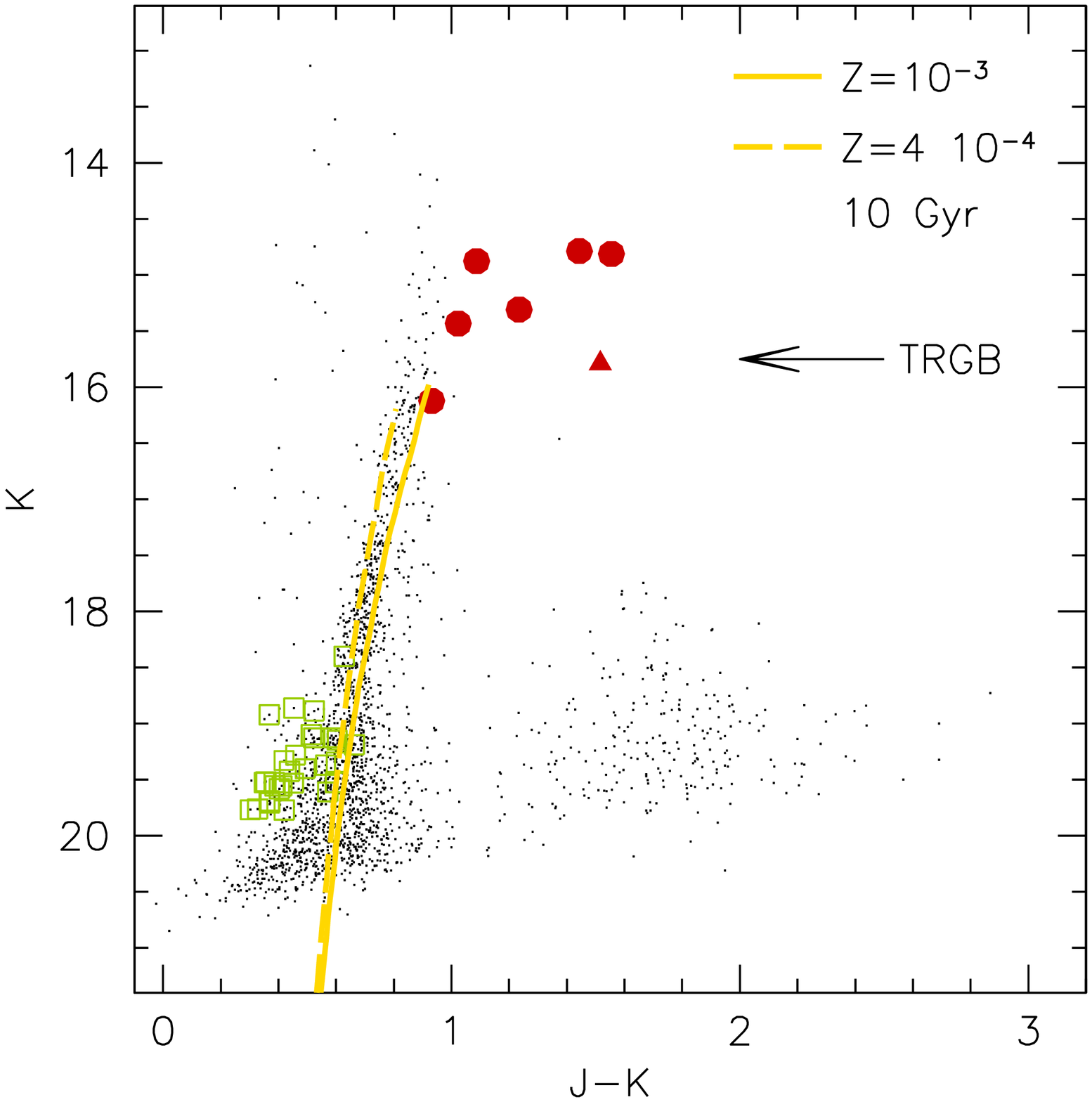}&
\includegraphics[clip,width=.45\textwidth]{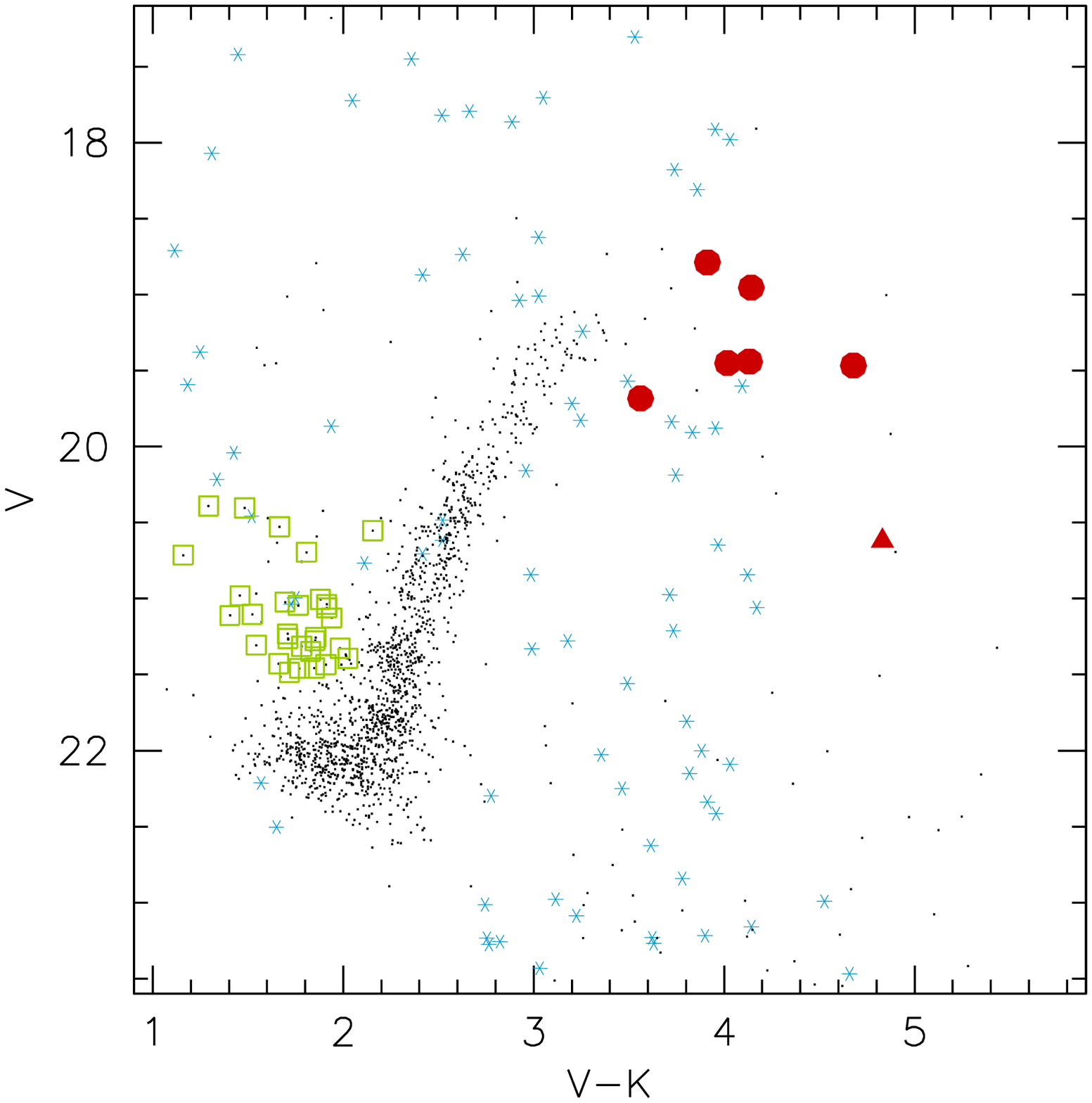}\\
\includegraphics[clip,width=.45\textwidth]{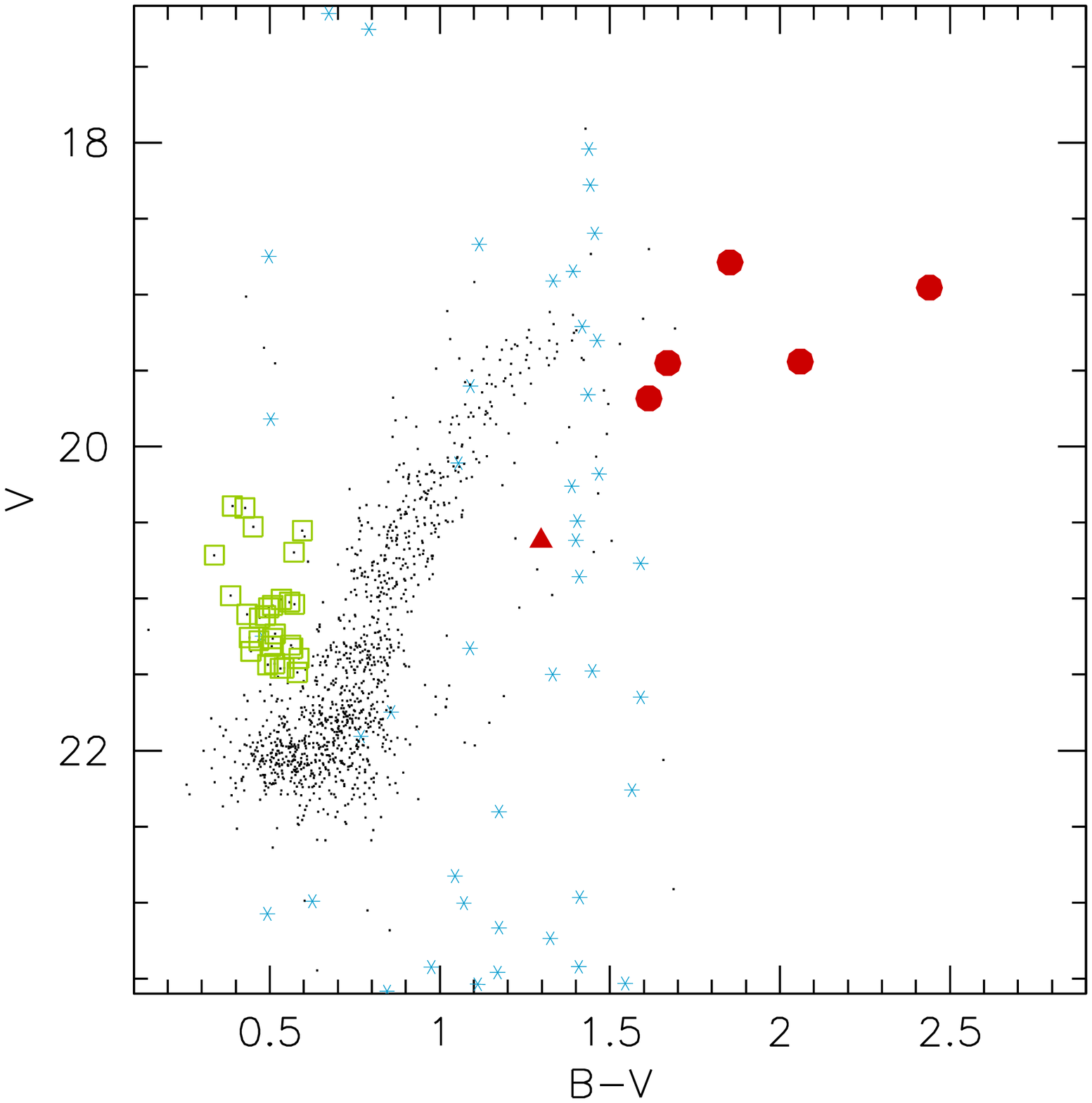}&
\includegraphics[clip,width=.45\textwidth]{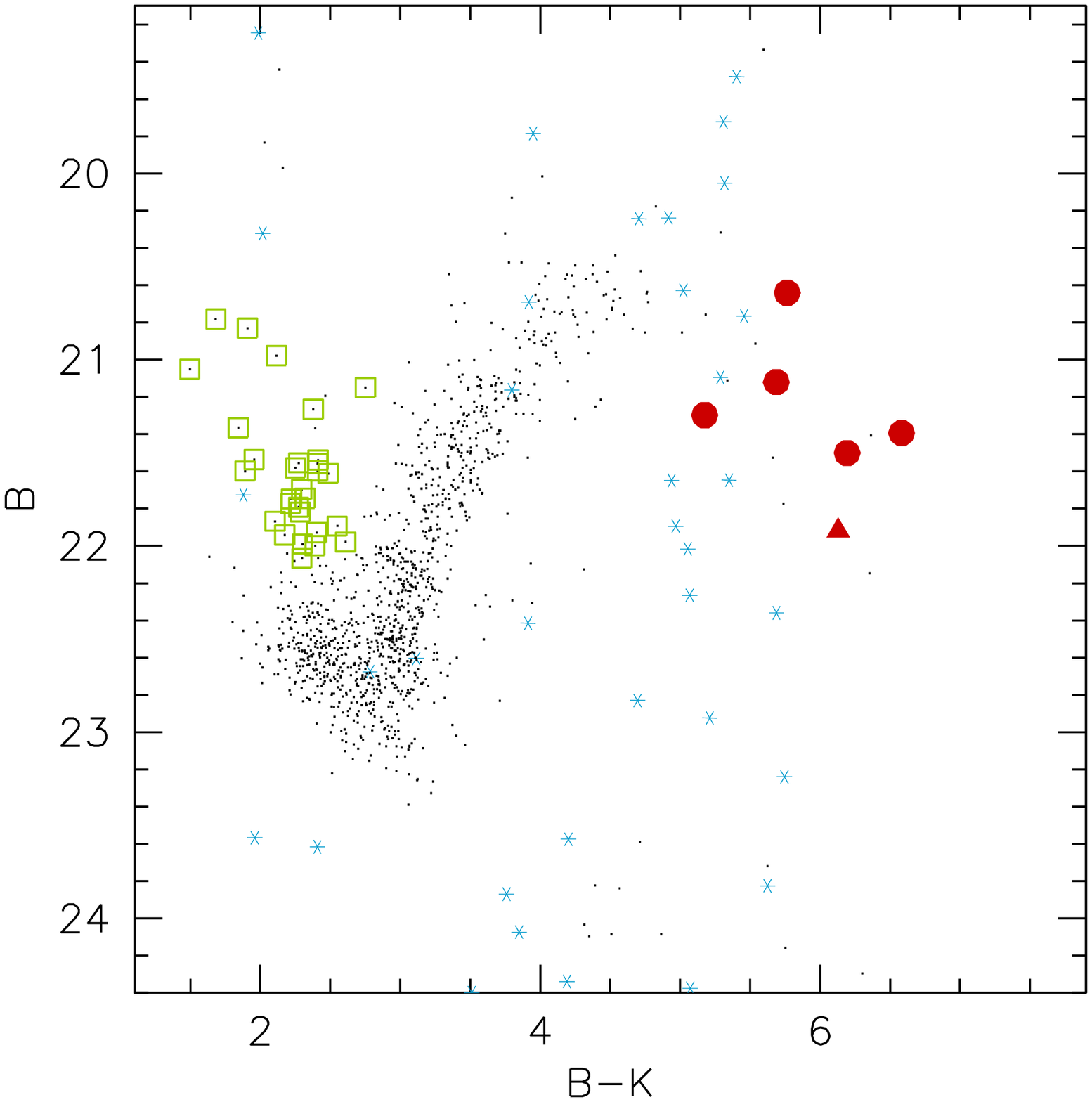}\\
\end{tabular}
\end{center}
  \caption{
  Colour-magnitude diagrams of \leoii\ from NIR and combined
  optical-NIR photometry.  In all diagrams, C stars
  spectroscopically identified by \citet{azzo+1985} are shown as {\it
  filled circles}, while their C star uncertain candidate  
  is marked by a {\it triangle}.  Superimposed on the
  \jks, $K_s$ CMD are theoretical isochrones from
  \citet{gira+2002}, for an age 10 Gyr and two metallicities close to
  the metallicity of \leoii, $Z=0.001$ and $Z=0.0004$.  In all except
  the \jks, $K_s$ CMD, we also show as {\it asterisks} the
  predicted Galactic foreground stars towards \leoii\ from a {\sc
  trilegal} simulation \citep{gira+2005}.  
  The He-burning
  stars on the ``yellow plume'', selected from the optical \bv,
  $V$ diagram, are shown as {\it open squares}.  They are best seen in
  the \bks, $B$ diagram, whereas the same stars are confused with
  the RGB in the \jks, $K_s$ CMD.
  \label{f:4cmd} 
}
\end{figure*}

Figure~\ref{f:4cmd} presents our \jks, $K_s$ colour-magnitude diagram of
\leoii, along with optical--NIR CMDs and an optical \bv, $V$ CMD.  A
selection based on the SHARP parameter was applied to remove noise
peaks, diffuse objects, and other spurious detections 
\referee{\citep[see, e.g.,][]{gull+2007sag}}.  The
optical--NIR CMDs and the optical CMD were obtained by adding data
obtained with the EMMI camera at the NTT at ESO/La Silla
\citep[][\sfhrizzip]{momany07}. Since the optical photometry refers to a
smaller central area ($9\farcm1 \times 9\farcm1$ in $V$, $6\farcm2 \times
6\farcm2$ in $B$), all diagrams involving optical data refer to a subset
of our WFCAM catalogue.

Our NIR CMD of \leoii\ shows a well populated RGB, with an TRGB
clearly visible at $K_s\sim 16$. Two 10 Gyr old isochrones
\citep{gira+2002}, with metallicity $Z=0.001$ and $Z=0.0004$
(corresponding to \feh\ $\sim-1.3$ and \feh\ $\sim-1.7$), are
superimposed to the CMD.
The CMD almost reaches the level of the HB, identified with the tail of
faint stars with colours bluer than the RGB, at a magnitude which is 
consistent with  $V_\text{HB}=22.18\pm0.18$ measured by
\citet{mighrich1996}. However, these stars will not be further analysed
since they fall close to the detection limit in the CMD. 

The sequence of stars brighter than the TRGB ($K_s \sim 15$) is 
identified with upper-AGB stars belonging to an intermediate-age stellar
component.  These stars appear to coincide with the sample of C stars
identified by \citet{azzo+1985}.  Note that they are more luminous 
than the TRGB  only in the \jks, $K_s$ diagram, while they appear
progressively fainter in bluer photometric bands.  This C star
population is quite small, in agreement with the low star formation rate
of \leoii\ at intermediate ages.  The AGB population will
be further discussed in Sects.~\ref{s:leo22col} and~\ref{s:leo2teo}.
The candidate C star with uncertain classification in \citet{azzo+1985}
does not share the spectral energy distribution of the C stars in the
different CMDs.  Visual inspection of our images confirms that it as a
background galaxy.

Note that both the prominent RGB sequence and the 
AGB component are absent in the CMD of an outer field of
\leoii, obtained from the detector No.~2 of WFCAM
(Fig.~\ref{f:leofield}).  This field is located at about 26\arcmin\ from
the centre of \leoii, corresponding to about 3 tidal radii.
A comparison with a simulation of the Galactic
foreground, obtained for the same field-of-view using the {\sc trilegal}
code \citep{gira+2005}, is shown in the right panel of
Fig.~\ref{f:leofield}.  The comparison indicates that the CMD of the
outer region mostly contains Milky Way stars 
(the vertical sequence with \jks~$\sim 0.8$ and bluer objects) 
and background galaxies (concentrated at \jks~$> 1$, $K_s > 18$). 

The wide baseline of optical--NIR diagrams provides the best
separation of some CMD features.  In Fig.~\ref{f:4cmd}, we plot as open
squares the stars on the ``yellow plume'' or ``vertical clump''
(\abbrev{VC}), just above the red clump, according to a magnitude and
colour selection from the optical \bv, $V$ diagram.  These stars are
best seen in the \bks, $B$ diagram as a vertical sequence originating
from \bks~$\sim 2.5$, $B \sim 22.5$ and extending up to $B \sim 21$;
while they are hardly detected in the \jks, $K_s$ NIR diagram,
an ambiguity that is explained by a combination of increasing
photometric errors at faint magnitudes and an intrinsically narrow
baseline.

The distribution of these ``vertical clump'' stars can be compared in
Fig.~\ref{f:4cmd} with the distribution of Milky Way foreground stars in
the direction of \leoii, obtained from the {\sc trilegal} code
\citep{gira+2005}. The projected Galactic contamination (plotted as
starred symbols) is insignificant in the yellow plume region of
the \bks, $B$ diagram, indicating that the vertical sequence is
certainly a \leoii\ stellar population.

In dwarf galaxies, this feature is usually attributed to a population
of core He burning stars a few hundred Myr to $\sim1$ Gyr old, which are
the descendants of stars located above the old main-sequence turnoff,
the so-called blue plume.  Thus, the detection of VC stars is generally
interpreted as evidence of recent star formation in these galaxies
\citep*[e.g., in Draco: ][]{aparicio01}.  
This is probably the case for \leoii, where there is evidence for an
increasing number of young stars toward the centre \citep{komi+2007}.
However, there is some evidence that the detection of a VC sequence in
dwarf spheroidal galaxies may not be sufficient to establish the
presence of recent star formation 
\citep{momany07,mape+2007}, since stars brighter than the HB have been
detected in globular clusters \citep[see the case of M80 in
][]{ferraro99}.
The agreement of the ``blue plume'' frequency in 7 dwarf spheroidal
galaxies (including \leoii) might be a hint that this population is
partly comprised of ``blue stragglers'', of which the VC population may
represent the evolved counterparts.

\realfigure{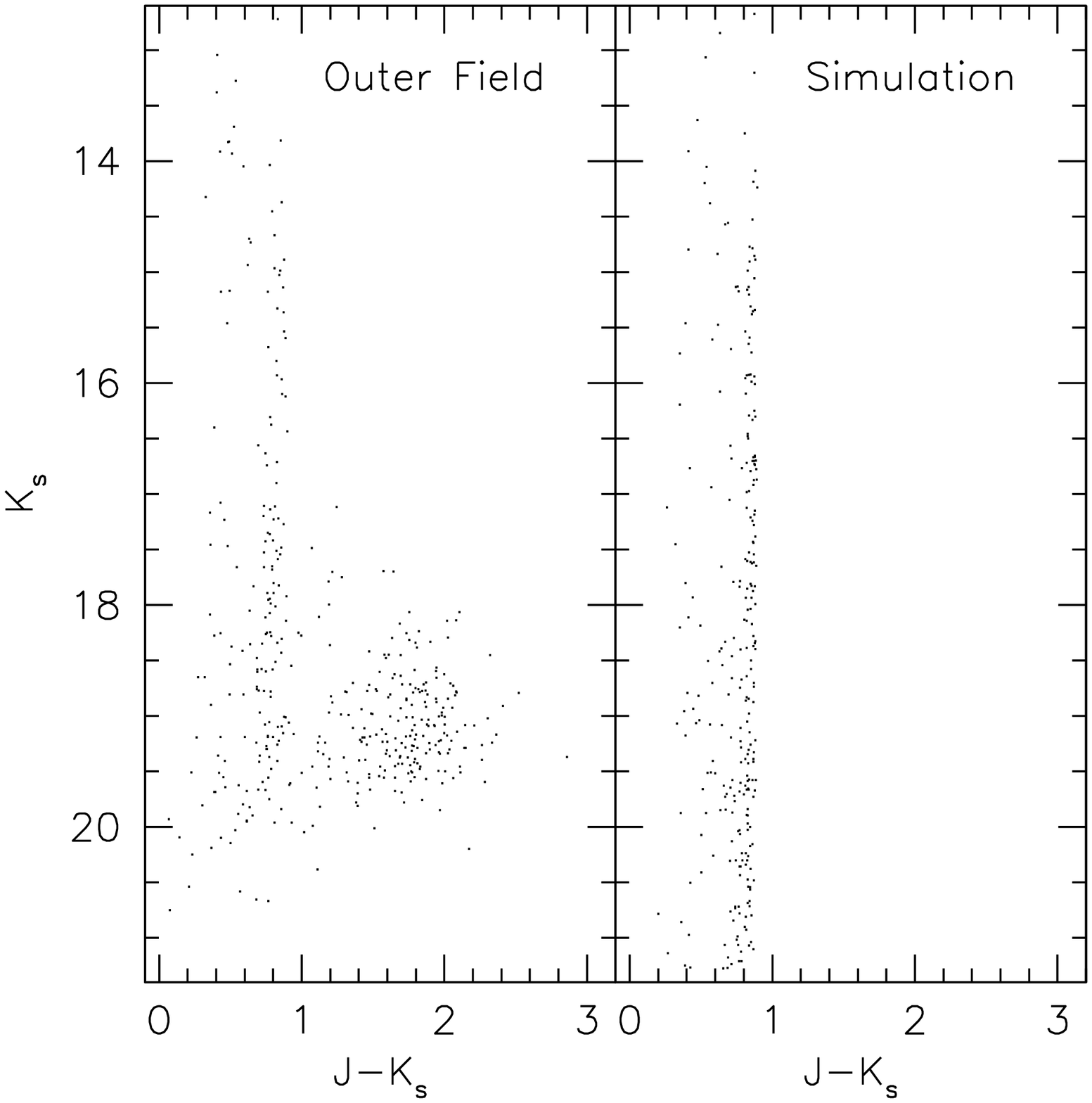}{
{\it Left:} 
colour-magnitude diagram of an outer field of \leoii, obtained from the
detector No.~2 of WFCAM, pointing at about 26\arcmin\ from the centre
of \leoii.  The vertical sequence having \jks~$ \la 0.8$ is consistent
with the expected Galactic foreground, while the group of objects redder
than \jks~$ = 1$ and fainter than $K_s = 18$ are unresolved
background galaxies.  {\it Right panel:} Milky Way stars in the \leoii\
line of sight as derived from a {\sc trilegal} simulation
\citep{gira+2005}, for the observed field-of-view, without photometric 
errors. }{f:leofield}

\section{Distance from the RGB tip}\label{s:leo2dist}

The luminosity of the TRGB in the $I$ band (where the dependence on the
age and chemical composition of the stellar population is at a minimum)
has long been used as a valuable standard candle
\citep{dacoarma1990,lee+1993}.
In the NIR, the TRGB luminosity depends on age and metallicity
in a more complex way. For instance, the $K_s$-band magnitude of
intermediate-age stars at the TRGB is fainter than that of old stars; 
while the TRGB $K_s$ luminosity rises with increasing metallicity
\citep[e.g.,][]{salagira2005}.  For a population that becomes more
metal-rich with time as a result of galaxy chemical evolution, the two
effects can partly balance.
In \citet{gull+2007for} we showed that, if the galaxy's SFH can be (even
roughly) estimated, quite accurate distance determinations can be
obtained from the TRGB at NIR wavelengths.

The method is applied here to measure the distance to \leoii\
independently of optical measurements. This will also be a useful test
of the reliability of distance estimates in the NIR domain, which is
important for next-generation instruments operating mainly at
NIR wavelengths (e.g., JWST, adaptive optics at Extremely Large
Telescopes).

We estimated the magnitude of the TRGB of \leoii\ by fitting its
$K_s$-band luminosity function to a step function convolved with a
Gaussian kernel representative of the photometric errors.  This method,
extensively applied by our group \citep[e.g.,][]{moma+2002}, was found
to give consistent results within $1 \sigma$ with the Maximum Likelihood
Algorithm of \citet{maka+2006} \citep[see][]{rizz+2007for}. 
The resulting $J$, $H$, and $K_s$ magnitudes of the TRGB are given in
Table~\ref{t:leo2trgb}.  The errors associated to these magnitudes are 
dominated by the uncertainty on the absolute photometric calibration,
because the error resulting from the TRGB fitting algorithm is less than
0.01 mag, and the internal photometric errors at the level of the TRGB
are negligible (see Sect.~\ref{s:leo2obs}).  The observed magnitude
were corrected for extinction using a reddening \ebv~$=0.03$ and the
\citet{rieklebo1985} reddening law.

To derive the distance to \leoii, the $JHK_s$ TRGB magnitudes were
compared with the empirical calibrations of $M_\lambda^{\rm TRGB}$ as a
function of \mh\ based on Galactic globular clusters \citep{vale+2004b},
whose intrinsic systematic error is $\sigma=0.16$ mag.
The adopted mean metallicity was \mh~$=-1.73$, in agreement with
\citet{koch+2006leo2} spectroscopy.  Given the relatively old age
distribution of \leoii, we found the population corrections 
\citep[calculated as in ][]{gull+2007for} to the TRGB 
magnitude to be negligible.

\mytab{
cccc}{
band&
$m^{\rm TRGB}$&
$(m-M)_0$&
$err_{(m-M)_0}$\\}{
$J$		&16.67	&21.73	&0.17\\
$H$		&15.90	&21.69	&0.19\\
$K_s$	&15.75	&21.58	&0.21\\}
{
Observed magnitude of the TRGB and distance moduli 
derived for \leoii\  from $JHK_s$ photometry. }{
t:leo2trgb}{
normalsize}

The distances derived from the $J$, $H$, and $K_s$ bands are also given
in Table~\ref{t:leo2trgb}.  The weighted mean is
$(m-M)_0=21.68\pm0.11$, 
\referee{in agreement with the value found by
\cite{lee1995} from the $I$ magnitude of the TRGB} and
intermediate between the ``short''
distance $(m-M)_0=21.55$ obtained by \citet{mighrich1996} from the $V$
magnitude of the HB and the ``long'' distance \dmo~$ = 21.84\pm0.13$
derived by \citet{bella+2005} from the $I$ magnitude of the TRGB.  
Therefore, our determination based on NIR magnitudes of the TRGB appears
to be consistent, within the errors, with the optical estimates.
The techniques explored in this paper will be useful to measure the
distance of stellar systems for which photometry of resolved stars will
become available only in the NIR.

\section{Metallicity}
\label{s:leo2metallicity}

\subsection{Metallicity distribution of RGB stars}

Given the uncertainties on the metallicity of \leoii\ and the
importance of the metallicity distribution function (\abbrev{MDF}) of
RGB stars as a constraint on models of chemical evolution of dwarf
galaxies, we used all the information from optical--NIR colours to
investigate the metallicities of stars in \leoii.
\referee{
A photometric MDF for stars in \leoii\ was derived following the
technique of \citet{savi+2000b}, extended to NIR wavelengths as
described in \citet{gull+2007for} for stars in Fornax dSph.
Stars on the upper RGB are interpolated in the \vks,
$M_K$ CMD across analytical fits to the RGB fiducial lines of template
Galactic globular clusters. The template clusters, spanning a wide
range in metallicity, are from \citet{vale+2004a}.
We used the global metallicity \mh, which measures the abundance of
all heavy elements. 
}
This parameter is the most appropriate to estimate the metallicities
of dwarf spheroidal galaxies (having [$\alpha$/Fe] ratios close to
solar) by comparison with the photometric properties of Milky Way
globular clusters, which generally show an overabundance of
$\alpha$-elements relative to iron that is a function of the cluster
metallicity \citep[see][ and refs.  therein]{geis+2007}.

\realfigurevert{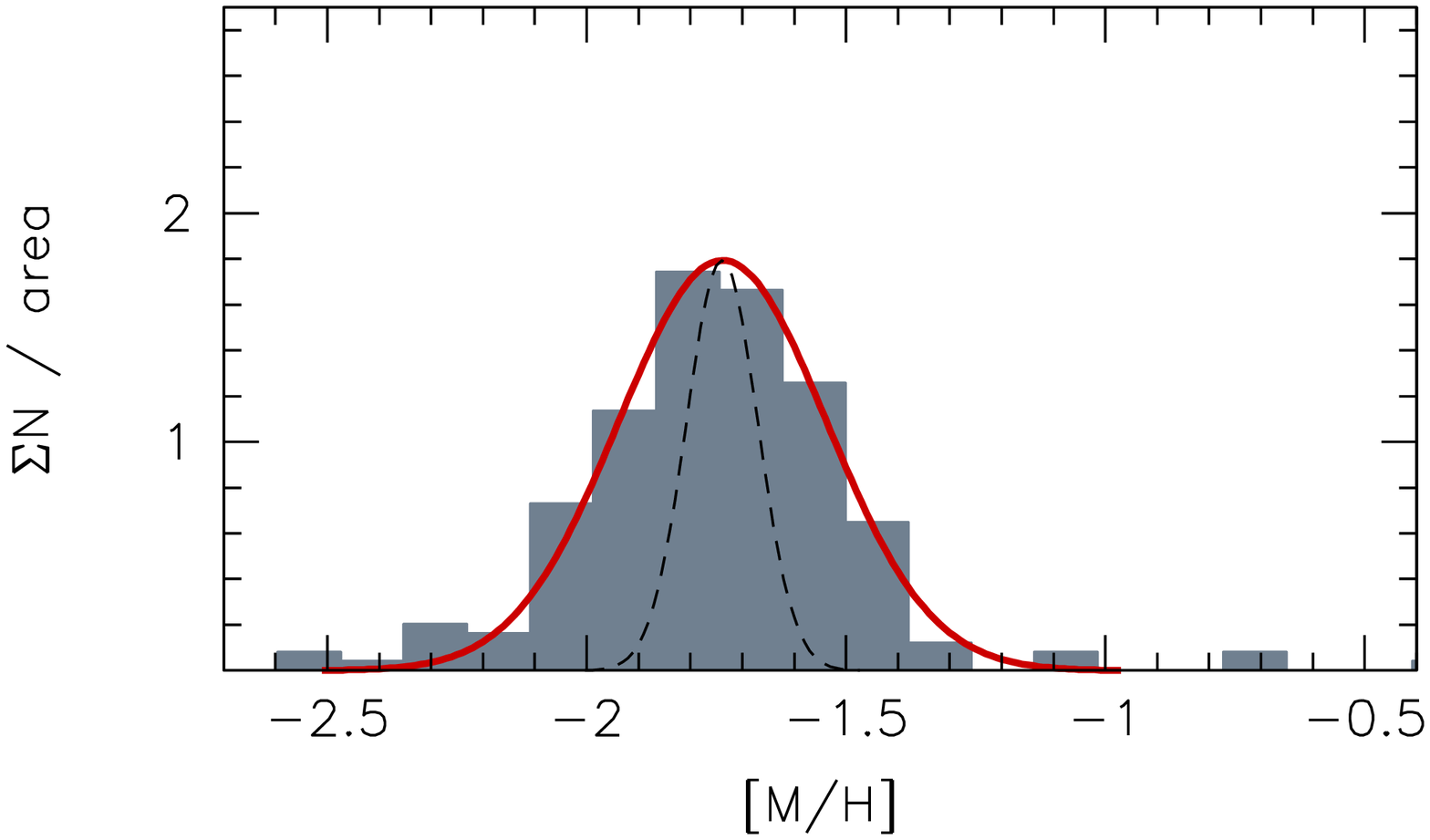}{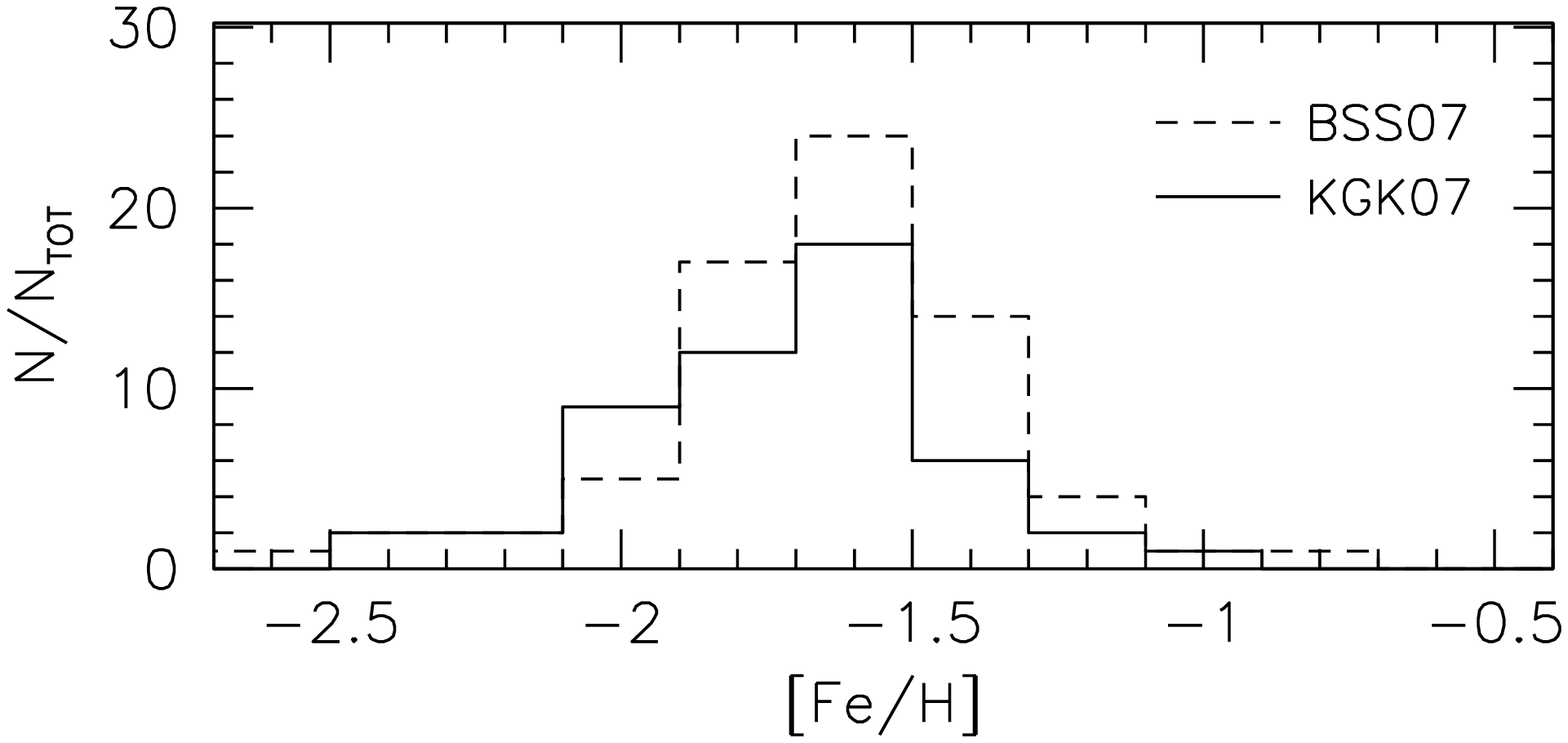}{ 
  {\it Upper panel: } the photometric Metallicity Distribution Function
  of RGB stars in \leoii, constructed using global metallicities \mh\
  derived from $(V-K_s)$ colours. The {\it solid line} is a Gaussian fit
  to the data, with mean \mh~$=-1.74$ and $\sigma_{\rm [M/H]}=$~0.20
  dex. The measurement scatter is represented by a Gaussian with
  $\sigma_{\text{instr}}=$~0.06 dex ({\it dashed curve}).  {\it Lower
  panel: } the spectroscopically derived MDFs of \citet{koch+2006leo2}
  ({\it solid line}) and \citet{bosl+2007} ({\it dashed histogram}),
  both on the \citet{carr+grat1997} scale.
}{f:leo2mdf}

The photometric MDF obtained for red giant stars in \leoii\ down to 2
mag below the TRGB, is shown in Fig.~\ref{f:leo2mdf} (upper panel).
The distribution is well described by a Gaussian function centred at
\mh\ $=-1.74$ and with a measured dispersion of 0.20 dex.  The internal
error in the same magnitude range was evaluated by applying the same
method to a synthetic CMD simulating a thin RGB, taking into account the
results of artificial star experiments.  The recovered metallicities
have a Gaussian distribution with a dispersion 0.06 mag, assumed to be
representative of the internal error of our metallicity measurements. By
quadratically subtracting this internal error from the measured width of
the MDF, we obtain a corrected dispersion 0.19 dex for the photometric
MDF of \leoii.

Our photometric MDF shown in Fig.~\ref{f:leo2mdf} is representative of
the {\it true} MDF only for stars as old as the Galactic GCs ($\sim
12.5$ Gyr).  However, a typical \leoii\ star is 9 Gyr old
\citep[e.g.][]{mighrich1996}, hence slightly bluer than globular cluster
stars of the same metallicity.  Therefore, as in \citet{gull+2007for},
we used theoretical isochrones to construct contours of constant \vks\
colours of RGB stars, as a function of both stellar age and metallicity,
and correct the measured metallicity for the age effect.
This is done by estimating the metallicity of a 9 Gyr old star having
the same colour as a 12.5 Gyr star with \mh~$=-1.74$, i.e. the mean of
the MDF. This differential approach overcomes any possible problems
with the absolute calibration of the isochrone colours.
With this assumption, the age correction results to be $\Delta$\mh\
$=0.10$, which is very small and comparable with the absolute
uncertainty of our method. Our major simplification is that {\it all}
stars have the same age, but it is adequate to calculate the {\it mean}
metallicity of \leoii\ stars.  By applying this correction, the mean
metallicity of \leoii\ turns out to be 
\mh~$=-1.64 \pm 0.06$ (random) $ \pm 0.17$ (systematic). 
\referee{ The systematic error was estimated from the uncertainty on
  the photometric zero point of our calibration, which is 0.07 mag on
  the \vks\ colour (the quadratic sum of the $V$ and $K_s$ zero-point
  uncertainties).  Shifting the \leoii\ RGB by $\pm 0.07$ mag in
  colour results in a $\pm 0.17$ dex variation in metallicity.  }

\subsection{Comparison with spectroscopy}

The derived metallicity is in excellent agreement with the two recent
spectroscopic results \feh~$=-1.73$ and $=-1.59$ by
\citet{koch+2006leo2} and \citet{bosl+2007}, respectively.
Our metallicity distribution of \leoii\ RGB stars, as inferred from
\vks\ colours, is compared with the spectroscopic metallicity
distributions from \citet{bosl+2007} and \citet{koch+2006leo2} in
Fig.~\ref{f:leo2mdf} (lower panel).  The distributions are basically
consistent, except for a slightly lower mean metal abundance from
photometry.  In particular, the range in metallicity (\abbrev{FWHM} of
the distributions) is comparable.

\realfigure{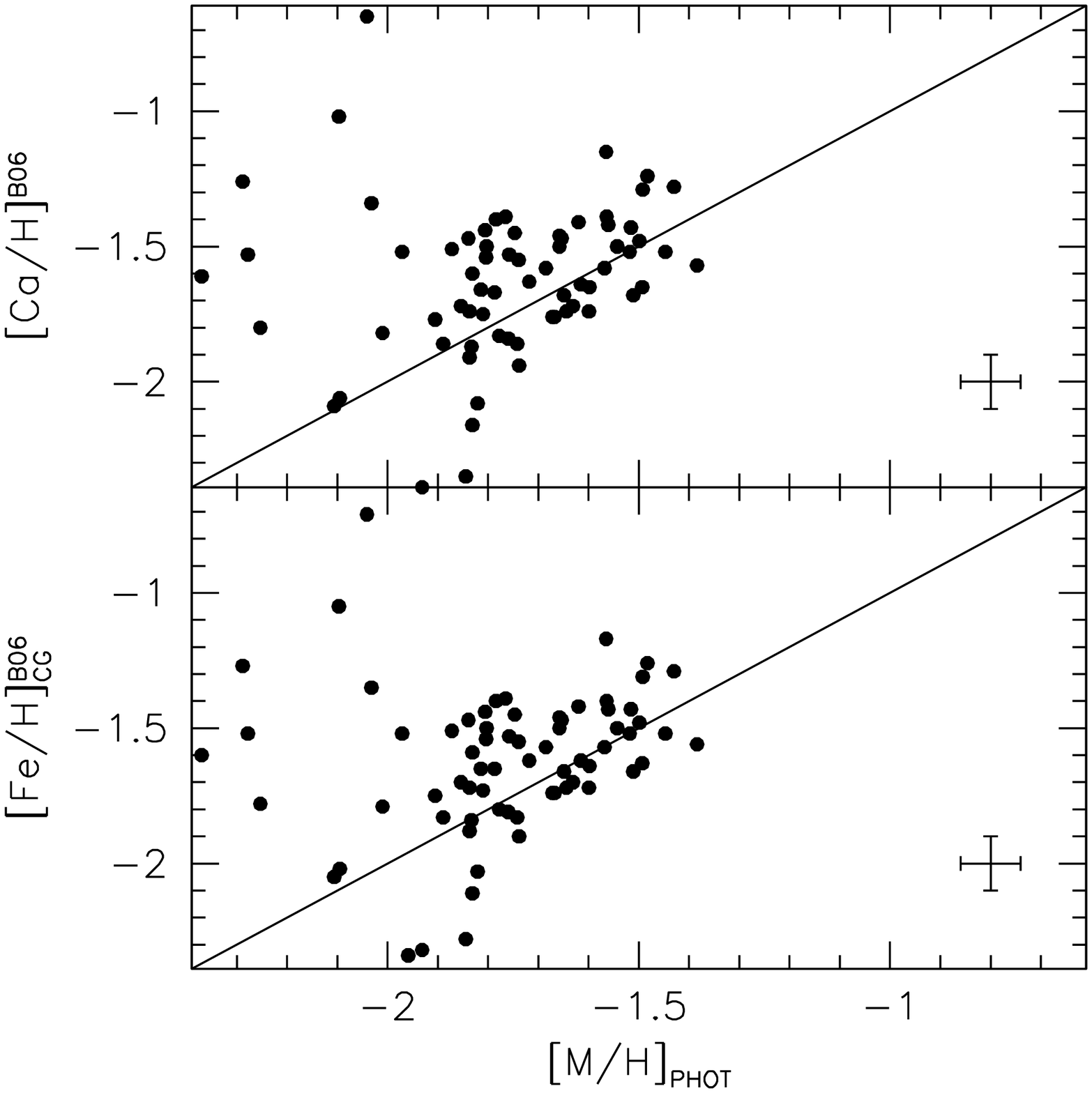}{Comparison of our measured
  photometric metallicities with the spectroscopic results of
  \citet{bosl+2007}.  In the {\it upper panel} we plot the spectroscopic
  data on the [Ca/H] scale of the authors, while the calibration based
  on \feh\ is shown in the {\it lower panel}. The error crosses
  represent the mean uncertainties of spectroscopic metallicities, and
  the $1 \sigma$ internal errors of our photometric determinations (not
  including age effects and systematic uncertainties).}{f:bosl}

The significant overlap between our sample of RGB stars and the catalogues of
\citet{bosl+2007} and \citet{koch+2006leo2} allows us a direct
comparison of photometric and spectroscopic metallicity estimates on a
star-by-star basis.

Figure~\ref{f:bosl} shows a comparison with the results of
\citet{bosl+2007}, for 71 stars in common with our sample.  In the
metallicity range typical of \leoii\ stars, the two calibrations adopted
by those authors (as a function of \feh\ and \cah) yield
\feh~$\simeq$~\cah.
Indeed, the relations presented in Fig.~\ref{f:bosl} for the two scales
are quite similar. For both scales, the overall agreement between
spectroscopic and photometric metallicities appears to be good.  The
most noteworthy difference is for the bluest stars, whose spread is
higher and spectroscopic metallicities are systematically higher.
A possible explanation is that the photometric metallicities are
underestimated because some stars are actually younger (hence bluer)
than old RGB stars. One alternative possibility is that calibration
uncertainties and internal errors affect in some way the spectroscopic
measurements at low metallicity.

\realfigure{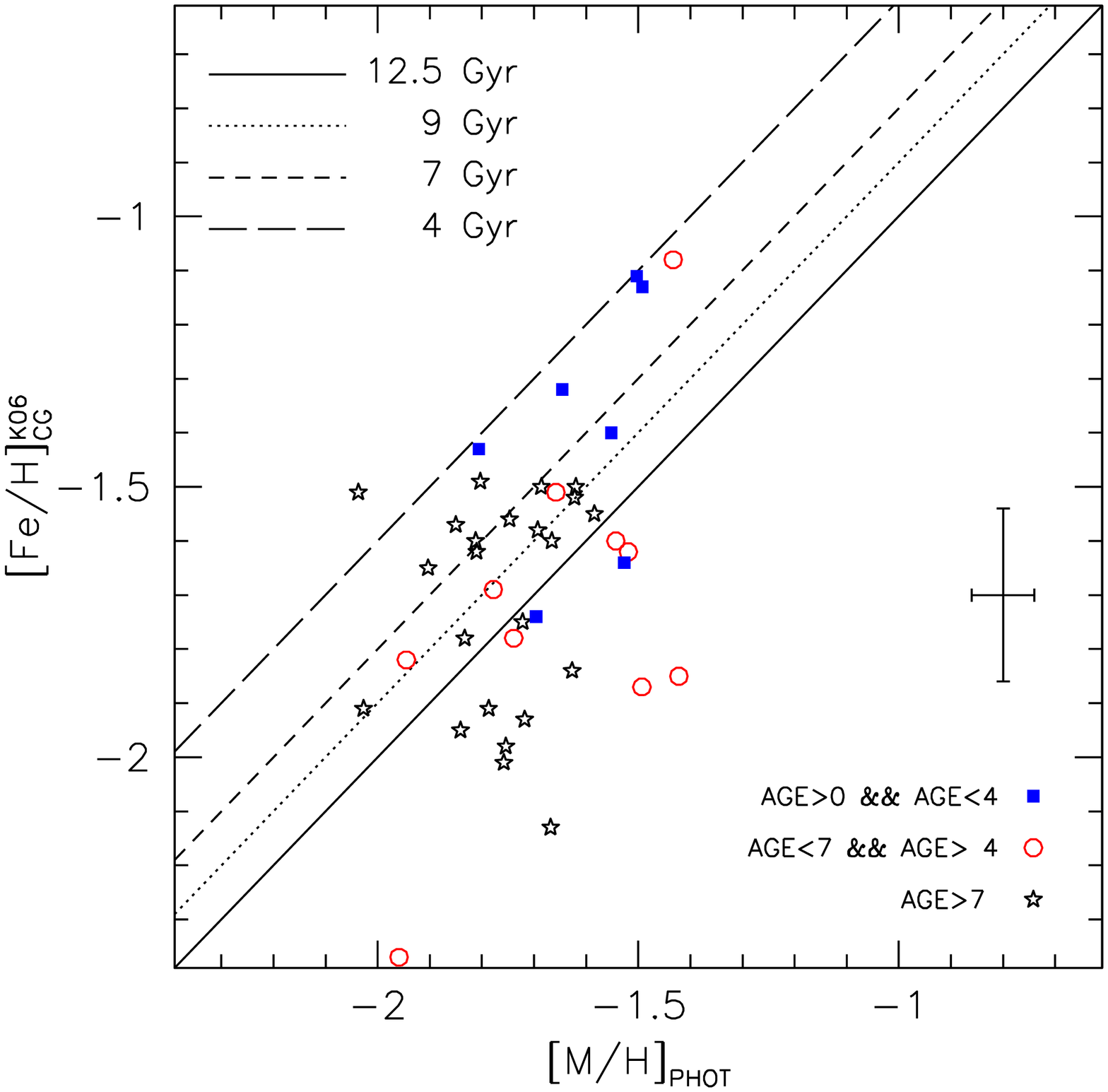}{Comparison of individual stellar
  metallicities with those derived by \citet{koch+2006leo2}, using
  different symbols for stars in different age intervals, as estimated
  by Koch and coll.  The lines represent the expected relations when the
  effects of age on \vks\ colours are taken into account.  For example, 4
  Gyr old stars are expected to follow the long-dashed line. Error bars
  as in Fig.~\ref{f:bosl}. 
}{f:kochstars}

\realfiguretwofig{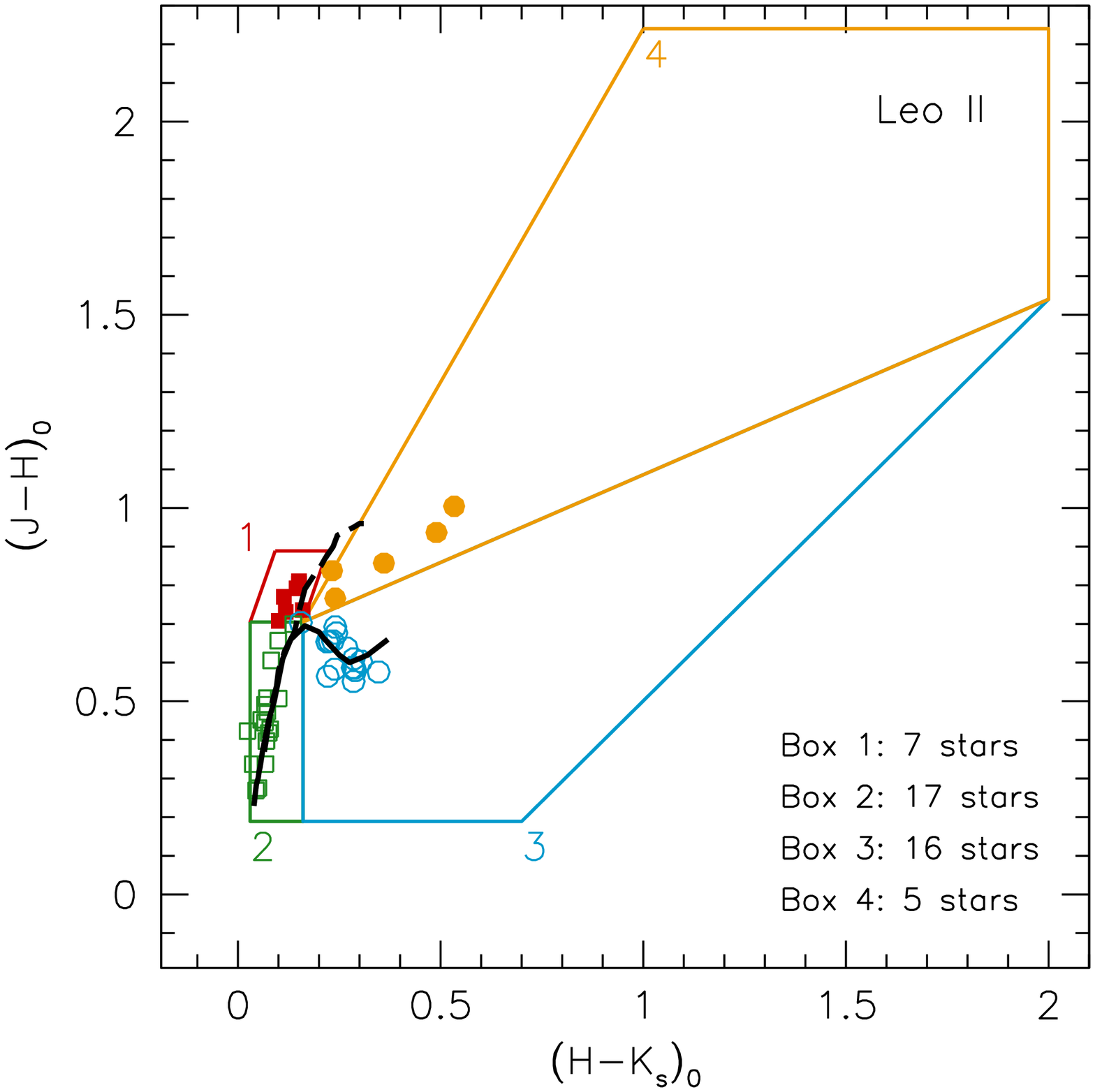}{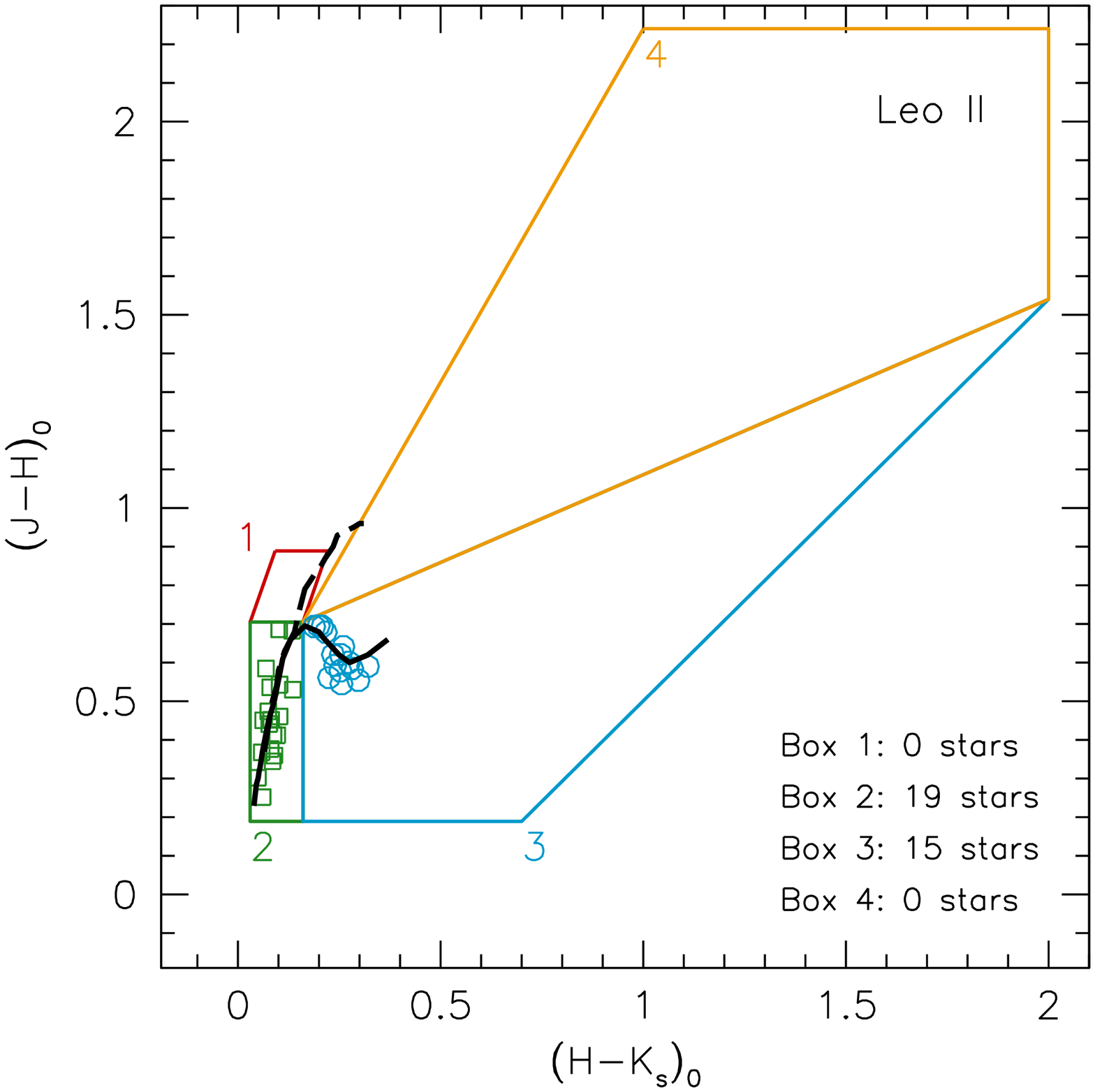} { {\it
  Left Panel:} the two-colour diagram of \leoii\ stars brighter than the
  TRGB ($K_s=15.75$), with superimposed the regions we have used to
  discriminate stars in \leoii\ from those in the Milky Way: region (1)
  are probable \leoii\ O-rich AGB stars, regions (2) and (3) are dwarf
  Galactic stars, region (4) is populated by C stars in \leoii.
  Different symbols indicate stars in different regions.
  The loci of giant stars and
  main-sequence dwarf stars \citep[from][]{bessbret1988} are shown as a
  {\it dashed} and {\it solid line}, respectively.
{\it Right Panel:} the same, for the outer field. Note the absence of
stars belonging to \leoii. 
}{f:2colbox}

A comparison of our photometric metallicities (with no age correction
applied) with the spectroscopic results of \citet{koch+2006leo2} is
presented in Fig.~\ref{f:kochstars}.  The 41 stars in common with our
sample are divided in 3 age intervals, using the age estimates published
by \citet{koch+2006leo2}.
The stars with ages \citep[as derived by][]{koch+2006leo2} greater than
7 Gyr, and most of the stars with ages in the range 4 to 7 Gyr, are
broadly consistent, with a large scatter, with the bisector in
Fig.~\ref{f:kochstars}, i.e.  compatible with the age of Galactic
globular clusters.  

We have also plotted the age-corrected relations (represented by
different lines in Fig.~\ref{f:kochstars}), by calculating the
metallicity shifts to be applied to our photometric measures for young
stellar populations.  Assuming an age of 9, 7, and 4 Gyr, the
corrections are $\Delta$\mh\ $=0.10$, $0.20$, and $0.41$, respectively.
Indeed, the location of stars younger than 4 Gyr seems consistent with
the expected relation for 4 Gyr old stars.
In general, however, we note a sizeable scatter, even considering stars
within each age bin, and the ages estimated by \citet{koch+2006leo2} do
not appear to be closely correlated with the age-corrected relations in
Fig.~\ref{f:kochstars}.
Overall, the number of young stars in \leoii\ derived by
\citet{koch+2006leo2} appears to be larger than suggested by SFH
reconstructions based on HST photometry
\citep[][\sfhrizzip]{hern+2000,dolp2002}, and the mean age of RGB stars
is younger.
We note that our age corrections, based on the larger baseline of
$V-K_s$ colours, may give more precise age ranking than the $g-i$ colour
used by \citet{koch+2006leo2}.

A direct comparison of spectroscopic metallicities, which would be
interesting to assess the accuracy of spectroscopic metallicities, is
not possible due to the absence of overlap between the samples of
\citet{koch+2006leo2} and \citet{bosl+2007}.

\section{Two-colour diagrams: selection of AGB stars}\label{s:leo22col}

In this section we present the two-colours diagram, which is used to
select AGB stars in \leoii.  This diagram is a powerful tool to separate
the foreground Milky Way stellar population, and allows a separation of
carbon and oxygen-rich stars \citep{aaro+moul1985,bessbret1988}.

Figure~\ref{f:2colbox} shows the NIR two-colour diagrams of
\leoii\ and the external field.  We selected only stars brighter than
the TRGB ($K_s=15.75$) to exclude RGB stars.  Stars are located in well
defined sequences, within the regions outlined in
Fig.~\ref{f:2colbox}.  All stars located in regions 2 and 3 are found
along the dwarf stars locus defined by \citet{bessbret1988}.
The number of stars in each region are also given in
Fig.~\ref{f:2colbox} for the field centred on \leoii\ and the external
field.  The number of stars in regions 2 and 3 are the same,
within statistical fluctuations, in the two fields.  We can therefore
conclude that {\it all} stars in regions 2 and 3 are Milky Way dwarfs.  
On the other hand, stars in region 1 and 4 are found only in
the field centred on \leoii, and we can conclude that they are all \leoii\
members. 
Being brighter than the TRGB, they can only be 
AGB stars. For stellar populations younger than those present in \leoii,
core He-burning red supergiants should be also considered.

All of the five stars in region 4 were identified as C stars by
\citet{azzo+1985}.  One more C star in their catalogue (ALW5) is
slightly fainter than the TRGB (see Fig.~\ref{f:4cmd}) 
and hence is not marked in Fig.~\ref{f:2colbox}. If plotted, it would
fall in region 3. 

\realfiguretwofig{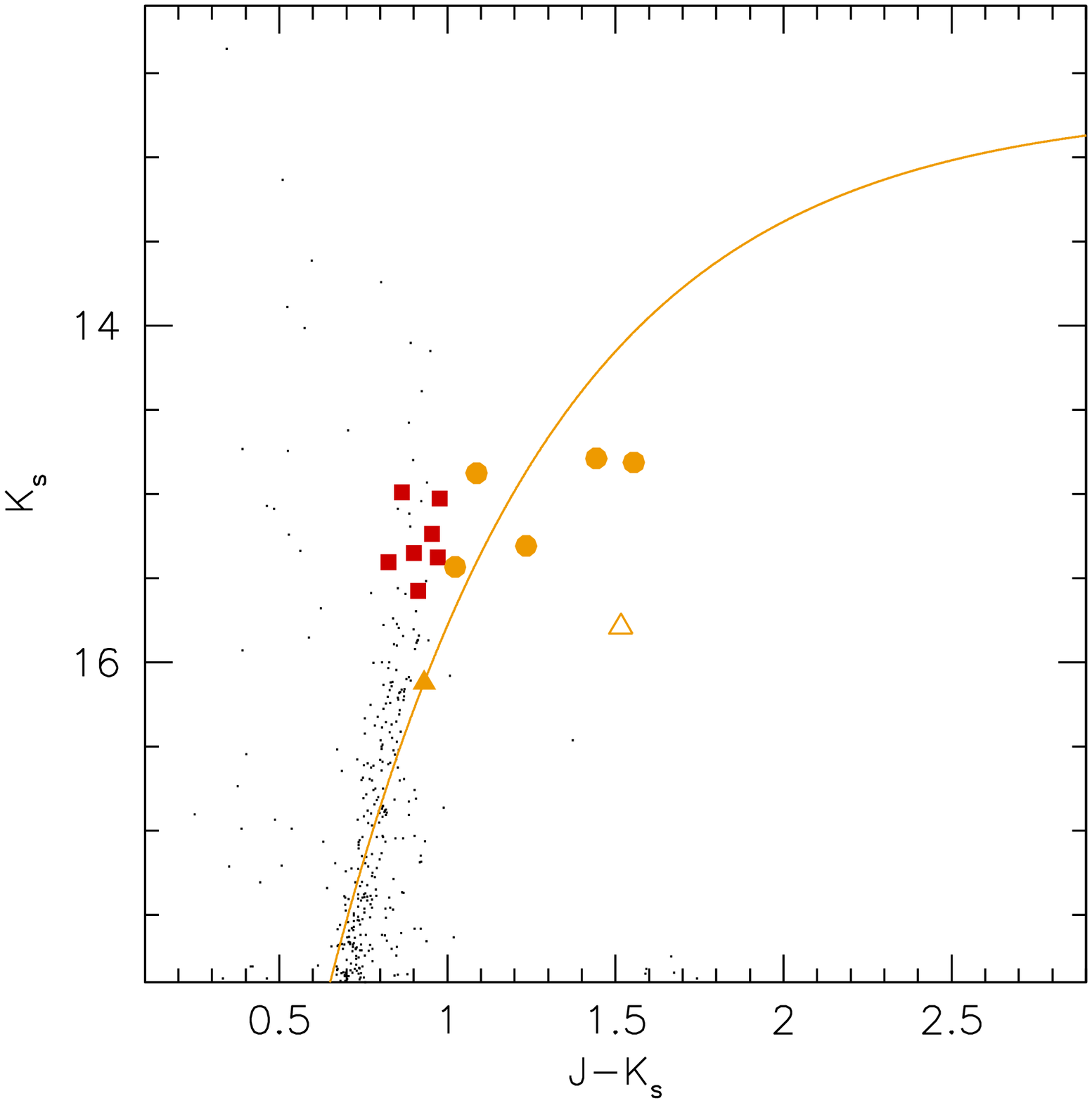}{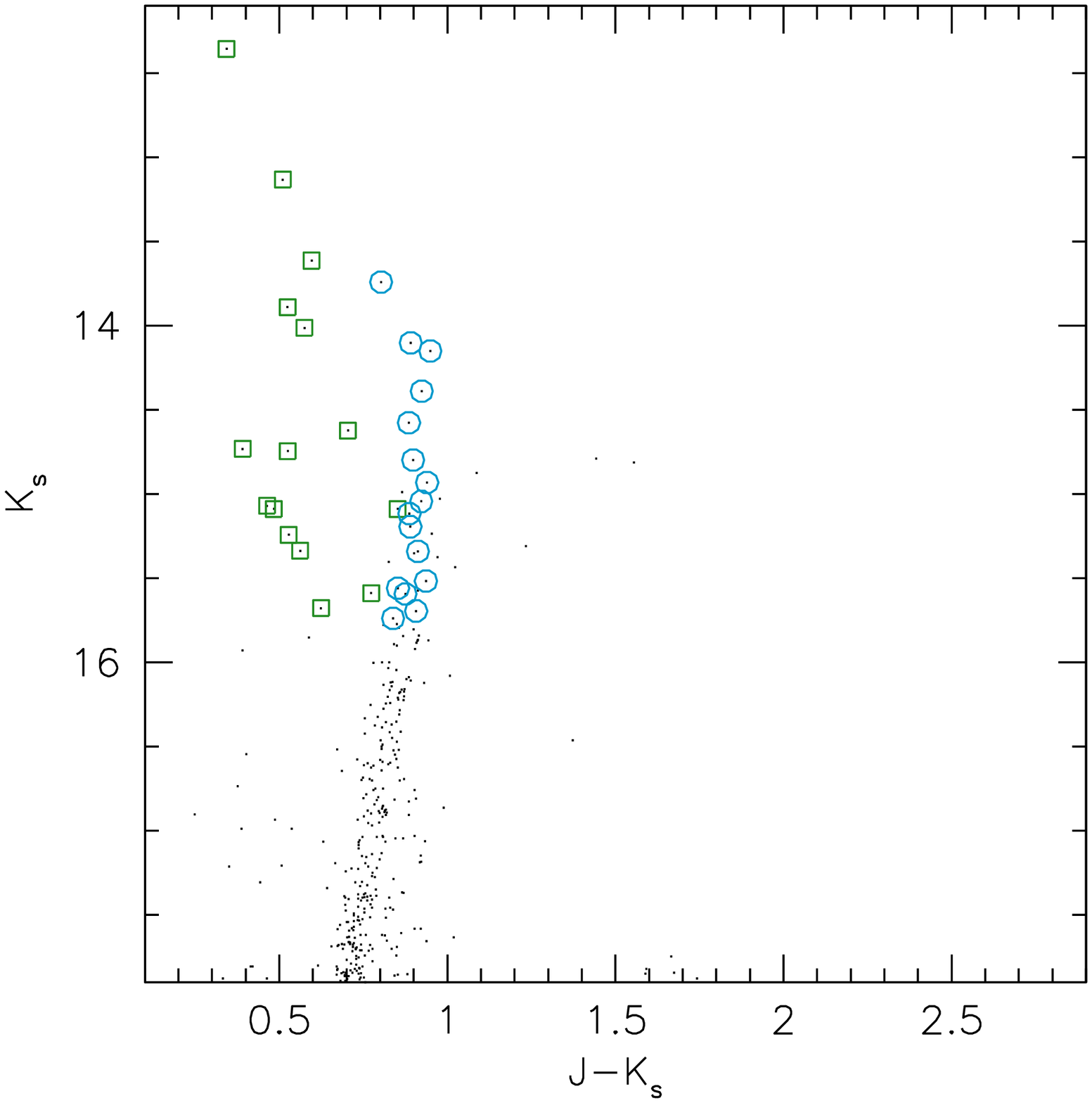}{
  The position of \leoii\ and foreground stars in the CMD according to
  the classification shown in Fig.~\ref{f:2colbox} for stars brighter
  than the \abbrev{TRGB}.
  {\it Left: } stars in the regions 1 ({\it squares}) and 4 ({\it filled
  circles}).  The {\it open triangle} represents ALW2, a misidentified
  background galaxy, while the {\it filled triangle} is a C star found
  just below the TRGB \citep[ALW5,][]{azzo+1985}.  The {\it solid curve}
  is the mean colour-magnitude relation for C star in Local Group dwarf
  galaxies \citep{tott+2000}, scaled to the distance of \leoii.
  {\it Right panel: } the same, for probable Milky Way stars in the
  regions 2 ({\it open squares}) and 3 ({\it open circles}) of the
  two-colour diagram.
}{f:cmdbox}

Figure~\ref{f:cmdbox} shows the location in the CMD of the stars
classified using the NIR two-colour diagram. Stars in the regions 1 and
4 are consistent with the expected loci of M and C stars respectively,
as judged from the location of AGB stars with spectroscopic
classification in the CMD of Fornax dSph \citep[see][and ref.s
therein]{gull+2007for}.
The C star population, in particular, agrees well with the mean
colour-magnitude relation for C stars in LG dwarf galaxies derived by
\citet{tott+2000}, scaled to the distance of \leoii\ discussed in
Sect.~\ref{s:leo2dist}.  NIR photometry of all C stars classified by
\citet{azzo+1985} and probable O-rich AGB stars selected by us in
region~1, is given in Table~\ref{t:agb}.

\mytabbig{
r c c c c c c c}{
\multicolumn{1}{c}{ID}&
\multicolumn{1}{c}{$\alpha$ (J2000)}&
\multicolumn{1}{c}{$\delta$ (J2000)}&
$J$&
$H$&
$K_s$&
type &
note\\}
{
 1661	& 11:13:12.82 &+22:11:14.1 &   16.231 &   15.284 &   14.788 &C &ALW1\\
 1671	& 11:13:20.64 &+22:11:16.3 &   16.366 &   15.351 &   14.812 &C &ALW3\\
 659	& 11:13:23.48 &+22:07:58.4 &   16.542 &   15.674 &   15.308 &C &ALW4\\
 828	& 11:13:23.97 &+22:08:29.3 &   17.054 &   16.365 &   16.122 &C &ALW5\\
 1089	& 11:13:29.39 &+22:09:14.2 &   15.963 &   15.114 &   14.876 &C &ALW6\\
 1032	& 11:13:31.78 &+22:09:06.1 &   16.456 &   15.679 &   15.433 &C &ALW7\\
 781	& 11:13:20.83 &+22:08:22.9 &   16.005 &   15.184 &   15.027 &O &\\
 1215	& 11:13:35.81 &+22:09:35.1 &   16.252 &   15.471 &   15.351 &O &\\
 1509	& 11:13:29.24 &+22:10:32.9 &   16.346 &   15.531 &   15.375 &O &\\
 1597	& 11:13:23.16 &+22:10:56.0 &   16.487 &   15.740 &   15.574 &O &\\
 1873	& 11:13:53.43 &+22:12:43.5 &   16.229 &   15.510 &   15.404 &O &\\
 1877	& 11:13:52.77 &+22:12:45.4 &   15.855 &   15.113 &   14.989 &O &\\
 1904	& 11:13:29.17 &+22:13:02.7 &   16.189 &   15.387 &   15.236 &O &\\
}
{NIR photometry of \leoii\ 
\referee{candidate O-rich and C-rich} AGB stars. 
The identifiers are those in our photometric catalogue. 
For the C stars, the names in \citet{azzo+1985} are also given.}
{t:agb}{normalsize}

In the following, we consider all the C stars identified by
\citet{azzo+1985}, including star ALW5, which is fainter than the TRGB
and was therefore not included in Fig.~\ref{f:2colbox}.  The objects
ALW2, misidentified by \citet{azzo+1985} as a C star, is not included in
our analysis.  All the remaining stars by \citet{azzo+1985} are
compatible with our C star selection and our observations cover all
\leoii, nearly out to the tidal radius.  We therefore conclude that the
complete population of \leoii\ C stars in the central $13\farcm6 \times
13\farcm6$ area covered by our observations, is formed by 6 objects
(excluding ALW2).  We finally note that \citet{azzo2000} stated they
found 2 new ones but without providing further details.  In our
selection we have no indications for the presence of other objects in
addition to the 7 discussed here.  Finally we consider as O-rich AGB
stars the 7 stars found in region 1 in Fig.~\ref{f:2colbox}. 

\section{Comparison with theoretical models}\label{s:leo2teo}

A distinctive feature of present observations is that they sample quite
completely the optically-visible AGB population of \leoii\ in the
surveyed area. The only AGB stars expected not to be present in our data
are those so strongly absorbed by circumstellar dust to become invisible
even in the NIR.

Such complete catalogues of AGB stars in nearby galaxies are rare. The
best such data are no doubt those for the LMC and SMC, fully sampled in
the $IJHK_{\rm s}$ bands of DENIS and 2MASS
\citep[see][]{cion+1999,nikowein2000} and now being sampled in the
mid-IR \citep[e.g][]{blum+2006, bola+2007}. Compared to these galaxies,
\leoii\ is more metal-poor (Sect.~\ref{s:leo2metallicity}), and presents
a much simpler history of star formation, concentrated at old ages.
These particularities provide us with a unique opportunity to test
present-day AGB models in the interval of low masses and low
metallicities.

\subsection{Simulating the photometry}

Having this goal in mind, we will try to fit the \leoii\
observed AGB population with the recent set of thermally pulsing AGB
(\abbrev{TP-AGB}) evolutionary tracks from \citet{marigira2007}.  Added
to the \citet{gira+2000} tracks for the pre-TP-AGB evolution, they are
converted to stellar isochrones as described in \citet{mari+2008} 
and fed to the {\sc trilegal} 
population synthesis code for simulating the
photometry of resolved stellar populations \citep[][{\tt
http://trilegal.kuleuven.be/}]{gira+2005}. Since the details of the
TP-AGB implementation in {\sc trilegal} are provided in separate papers
\citep[e.g.][and work in preparation]{giramari2007}, suffice here to
recall the basic aspects of the model simulations:
\begin{itemize} 
\item The Milky Way foreground is simulated as in \citet{gira+2005},
 including the main disk and halo components and for the same
area of our observations.
\item The \leoii\  galaxy is set at a distance of 205 Kpc (this paper).
Reddening is ignored since it is negligible.
\item 
We fix the size of the simulations by
reproducing the star counts in the upper 2 magnitudes of the RGB --
for which our observations are quite complete -- together with the
relative star formation rate (SFR).
\item The relative SFR of \leoii\ is taken from two different sources,
as depicted in the upper panel of Fig.~\ref{fig_sfr}: \citet{dolp+2005}
and \sfhrizzi.  In both cases the SFR is derived from the inversion of a
deep CMD from HST. Although the error bars in these SFR determinations
are quite significant, these SFRs  show that the bulk of star formation
in \leoii\ was confined to ages larger than 2 Gyr, and concentrated at
$>4$~Gyr.  More details are given later.
\item The metallicity is taken from the photometric 
determination of Sect.~\ref{s:leo2metallicity}, 
i.e. we use a Gaussian distribution of mean
[M/H]$=-1.64$ and dispersion 0.2~dex. This metallicity distribution is
assumed to be the same for all ages, which is likely a good
approximation since the age--metallicity relation derived by 
\citet[][their figure 11]{koch+2006leo2}
is essentially flat for ages above 5 Gyr.
\item Each simulated star is converted in the 2MASS system using 
an updated version of the \citet*{bona+2004} transformations. In
particular, the transformations for carbon stars are now derived from
\citet{loid+2001} spectra.  In
addition, using \citet{groe2006} tables we correct the photometry
for the effect of circumstellar dust in mass-losing AGB stars. 
The 60\% Silicate + 40\% AlOx and 85\% AMC + 15\% SiC dust mixtures 
are assumed for O-rich and C-rich stars, respectively.
\item For TP-AGB stars, the pulse cycle luminosity and \Teff\ 
variations are also simulated; long period variability is not. For the
high-amplitude Miras, variability may provide an additional scatter of
about 1~mag 
\citep[see, e.g., ][]{cion+2003} in the $K$ band. Our
simulations however predict that most of the AGB stars are first
overtone pulsators, which have much smaller amplitudes.
\item Photometric errors and completeness are simulated using the 
relations derived in Fig.~\ref{f:compl}. 
\item To reduce the statistic fluctuations in the numbers of 
predicted stars, each simulation is run at least 50 times with
different random seeds. When we refer to the ``expected numbers'' of
each kind of star, we are actually referring to the mean values and
standard deviations obtained from these many runs.
\end{itemize}

\realfigure{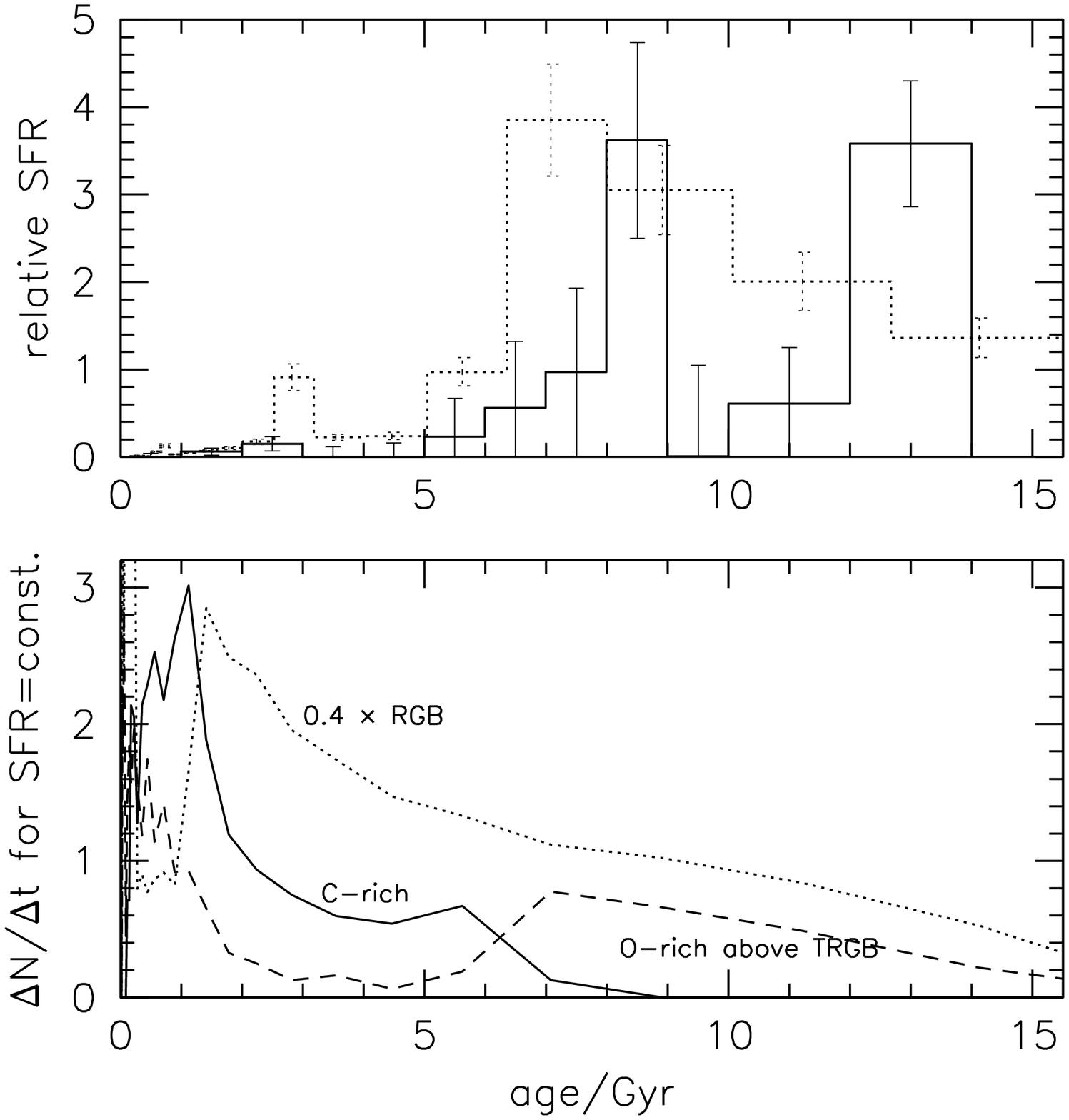}{ {\it Top panel:} The star-formation
  histories used in this work: \citet[{\it dotted line}]{dolp+2005} and
  \sfhrizzi\  ({\it solid line}) as a function of
  age. {\it Bottom panel:} The 
  production (number of stars per
  time interval) of different types of stars in our models as a function
  of age, for a galaxy model forming stars at a constant rate 
  (in mass per unit time) from 0 to
  15 Gyr, at a constant $Z=0.0004$ metallicity. The stellar kinds
  plotted are RGB stars within 2~mag of the TRGB (multiplied by 0.4;
  {\it dotted line}), and both O-rich giant stars above the TRGB
  ({\it dashed line}) and C-rich giants ({\it solid line}).}{fig_sfr}

Our simulations also take into account the error bars in
\citet{dolp+2005} and \sfhrizzi\ determinations of the SFH. These error
bars reflect both the intrinsic errors in the method of SFR-recovery,
and the small number statistics of the original HST data from which the
SFR is derived. For each age interval, we use a random SFR value 
drawn from a normal distribution centred at the mean SFR and with
the appropriate value of $\sigma$.
Since for some age bins the $1\sigma$ error bars are comparable to the
mean SFR, the negative values of SFR obtained at the youngest age bins
are set to zero. This produces a distribution of SFR values that is
non-symmetrical around the mean values, especially at the youngest age
bins.

One of such simulations is shown in Fig.~\ref{fig_simcmd}, which
resembles very much Fig.~\ref{f:4cmd}. First, a key to understand the
simulations is given by the \citet{mari+2008} isochrones for a few
selected ages. They show the location of the TP-AGB phase at quiescent
phases of H-shell burning, where these stars spend about 70~\% of
their life. The remaining $\sim30$~\% is spent at phases of lower
luminosity (and higher \Teff) after the occurrence of He-shell
pulses. The result is that the TP-AGB stars in the simulation are
typically found above the TRGB, in the same region defined by the
isochrones, but a tail of such objects (both C- and O-rich) extends
down to almost 2 mag below it. Moreover, a significant fraction of the
O-rich giants found above the TRGB are not genuine TP-AGB stars, but
either early-AGB stars or (more rarely, in the case of \leoii) 
core He burning stars belonging to the youngest populations.

Comparing the simulated data points with the observed ones of
Fig.~\ref{f:4cmd}, one notices that the main stellar features are
accounted for by the model. 
The simulations do not contain the objects observed at the bottom right
part of the diagram ($K_s>18$, \jks~$ > 1.2$), which likely correspond to
background galaxies \citep[e.g.][]{nikowein2000}.  Simulated
carbon stars are on average bluer than the observed ones, although 
a few very red dust-enshrouded objects are present in our simulations.
There are also 
minor offsets in colours of O-rich stars, that cannot be appreciated
in Fig.~\ref{fig_simcmd} because they are of the order of 0.05~mag.
It is interesting to note that 3 out of the 5 stars expected to be
C-rich, are located along the O-rich sequence, i.e., not exhibiting
redder \jks\  colours than those of the O-rich stars with the same
luminosity.  This prediction is consistent with the observed
colour-magnitude diagram presented in Fig.~\ref{f:4cmd} (top-left panel),
where 3 C-rich objects are just seen to lie on the O-rich giant branch.
On the theoretical ground this feature is explained considering that at
low metallicity the cooling effect on the stellar atmosphere due to the
carbon-enhanced molecular opacity becomes less efficient, as illustrated
by \citet[][their figure 7]{marigira2007}.

\realfigure{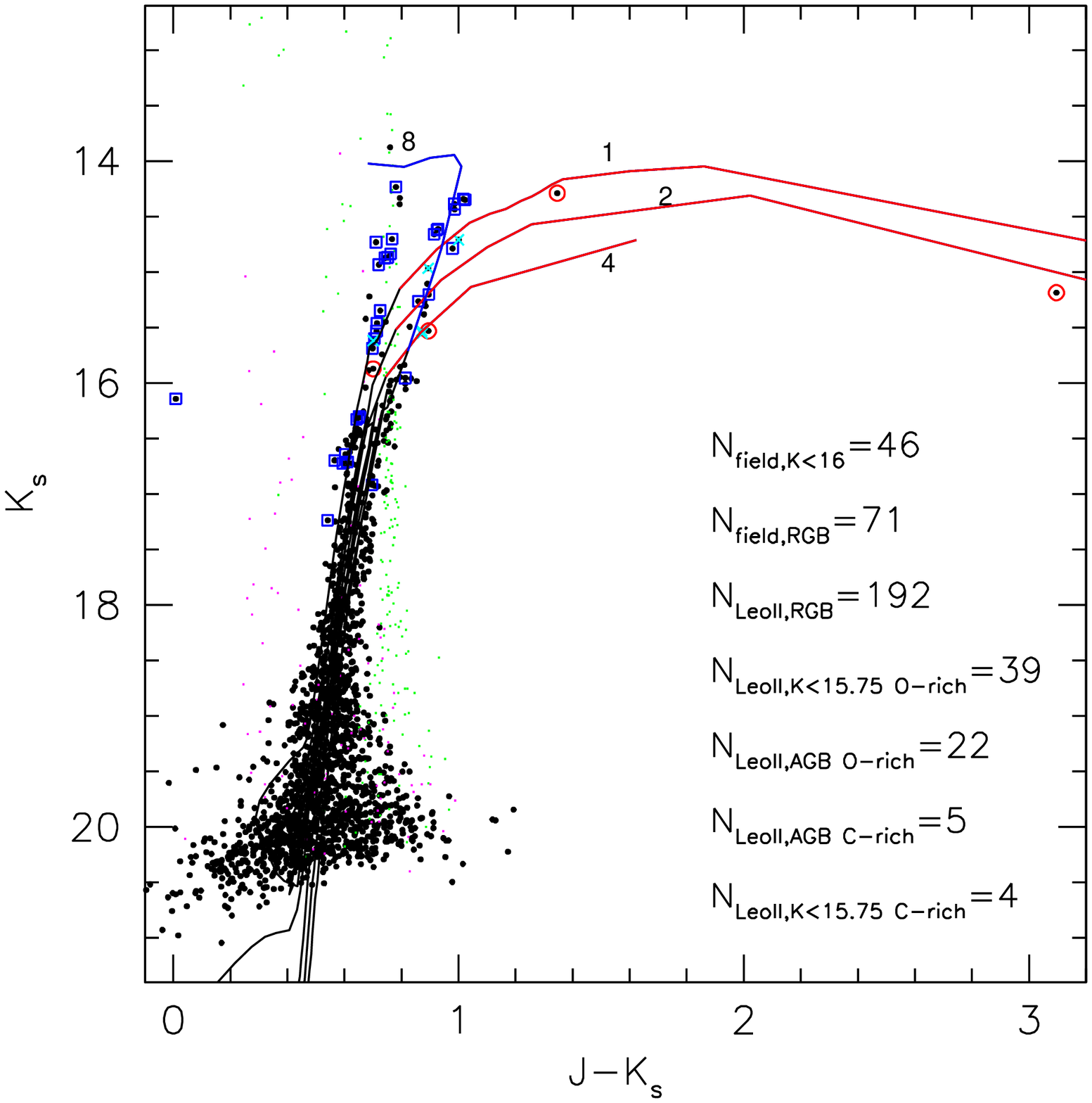}{ 
  An example of simulated 
  CMDs for \leoii, using the \citet{dolp+2005} SFH. In the electronic
  version of this paper, different colours mark different kinds of
  stars, namely: Milky Way disk (green crosses) and halo (magenta
  dots), \leoii\ pre-TP-AGB stars (dark dots),
  \leoii\ early-AGB stars above the TRGB magnitude (cyan crosses), 
  and \leoii\ TP-AGB stars both O-rich (blue
  squares) and C-rich (red circles). We also plot the \citet{mari+2008} 
  isochrones for $Z=0.0004$ and ages of 1, 2, 4, and 8 Gyr,
  shifted by the \leoii\ distance modulus. Dark lines mark phases
  previous to the TP-AGB, blue lines the O-rich TP-AGB, and red lines
  the C-rich one. The TP-AGB lines correspond to quiescent phases of
  evolution.
}{fig_simcmd}

\subsection{Comparing foreground and RGB counts}

We find that the expected number of foreground stars in our
0.052 deg$^2$ area, limited to the $13 < K_s < 16$ 
magnitude interval, is $38.6\pm6.8$; 
this is well compatible (within 1$\sigma$) with the 44
objects observed in regions 2 and 3 of Fig.~\ref{f:2colbox} 
at $K_s < 16$. This agreement is just
expected, since one of the deep fields used to calibrate {\sc trilegal} --
namely the CDFS \citep{groe+2002}  -- is located at the same
galactic longitude and at a similar latitude from the Galactic Plane
(i.e. $\ell=220\fdg0$, $b=-53\fdg9$) as \leoii\footnote{Note that
number counts in NIR bands are very much symmetrical with respect
to the Galactic Plane, at least for $|b|\ga20$~deg.}. Therefore, we
would expect that the typical errors in the predicted number counts at
the position of \leoii\  ($\ell=220\fdg2$, $b=+67\fdg2$) are similar to
those of the CDFS, i.e. of just $\sim10$~\% down to $K_s\sim18$
\citep[see Fig. 6 in][]{gira+2005}.

As the simulations predict the correct number of foreground stars at
$13 < K_s < 16$, they can also be used to infer the field contamination of
other CMD regions. We find that a total of $62.0\pm7.2$ foreground 
stars are expected to contaminate the uppermost 2~mag of the RGB. The
total observed number is 254.  Therefore, \leoii\  genuine RGB stars are
expected to be about 192 and outnumber the foreground contaminants in
the upper part of the RGB by a factor of about 4.  If we assume the
possible errors in the number of simulated foreground stars to be of
the order of 20 percent, they will have just modest consequences
(errors of the order of 8 \%) in determining the total \leoii\ 
mass to be simulated.  

\subsection{Comparing AGB counts}

We proceed with \leoii\  simulations with a total mass scaled such as
that $192\pm14$ RGB stars are produced in the upper 2 magnitudes of
the RGB. The results concerning the AGB are as follows.

For the \sfhrizzi\  SFR, we expect to find
$43.9\pm6.6$ O-rich giants above the TRGB (with $29.4\pm5.5$ being
genuine TP-AGB stars). The C-rich AGB stars are $4.4\pm2.4$ (with
$3.4\pm2.2$ above the TP-AGB). In comparison, the \leoii\ data presents 7
and 6 of such stars, respectively. There is a clear excess of O-rich
giants in the simulations, by a factor of about 6, which is extremely
unlikely to be due to statistical fluctuations. For the C stars,
instead, getting 6 stars out of an expected number of 4.4 is well inside
the 67~\% confidence level (CL) of a Poisson distribution.

Using the \citet{dolp+2005} SFH, $42.0\pm5.5$ O-rich giants are
predicted above the TRGB ($26.4\pm4.7$ genuine O-rich TP-AGB ones),
and $8.3\pm2.7$ C-rich ($5.7\pm2.1$ above the TAGB). The excess of
O-rich giants is again of a factor of about 6. For the C stars, the 8.3
predicted stars are again inside the 67~\% CL of a Poisson distribution
of the 6 observed ones. Therefore, also in this case the observed
C stars are compatible with the model predictions.

To understand why models using Rizzi's et al. SFH present about half of
the C stars as compared to the \citet{dolp+2005} case, it is instructive
to compare the two panels of Fig.~\ref{fig_sfr}, the top panel showing
the relative SFRs and the bottom one showing the age distribution of
different stars for a model galaxy of the same metallicity but with
constant SFR.
The bottom panel shows that the maximum age for the formation of carbon
stars, $t_{\rm C}^{\rm max}$, is close to 6 Gyr ago. This limit is
actually determined by the lifetime of the least massive TP-AGB 
stars in the \citet{marigira2007} models to experience the third
dredge up events, with 1.0~\Msun\ (6 Gyr).
Below $t_{\rm C}^{\rm max}$ the C stars predominate,
above it they are simply absent and the relative number of O-rich TP-AGB
stars increases. 
A substantial fraction of the star formation in \leoii\ has occurred close
to $t_{\rm C}^{\rm max}$, and this determines a marked dependence of the
C star counts on the details of the SFR at this age interval. Since the
\citet{dolp+2005} SFR presents a marked episode of star formation
between 6 and 8 Gyr, this determines the large number of C stars of the
corresponding model.

Were the mass limit for the dredge-up
to occur just 10~\% (or 0.1~\Msun) different, $t_{\rm C}^{\rm max}$
would change by as much as $\sim40$~\%.  This would impact
very much on the predicted numbers of C-type stars. Also the numbers
of O-rich AGB stars would be affected, although somewhat less, since
they reflect the complete SFR up to ages of 15~Gyr. The simple fact
that the numbers of predicted C-type AGB stars turn out to be
consistent with observations within the 95~\% CL of a Poisson
distribution, would be indicating that {\em the minimum mass for the
formation of C stars at low metallicities is indeed close to
1.0~\Msun.} This is an important indication for the theoretical
modeling of AGB stars. We recall that at LMC metallicities, the same
mass limit is closer to 1.4~\Msun. Classical models of stellar
evolution\footnote{For classical models we mean models following the
Schwarzschild criterion for convective borders, and solving separately
for the stellar structure and nucleosynthesis.}  have a strong
difficulty in reproducing such low mass values for the progenitors of
carbon stars \citep[see e.g.][]{herw2005,stan+2005}.

\subsection{Luminosity functions}

\realfigure{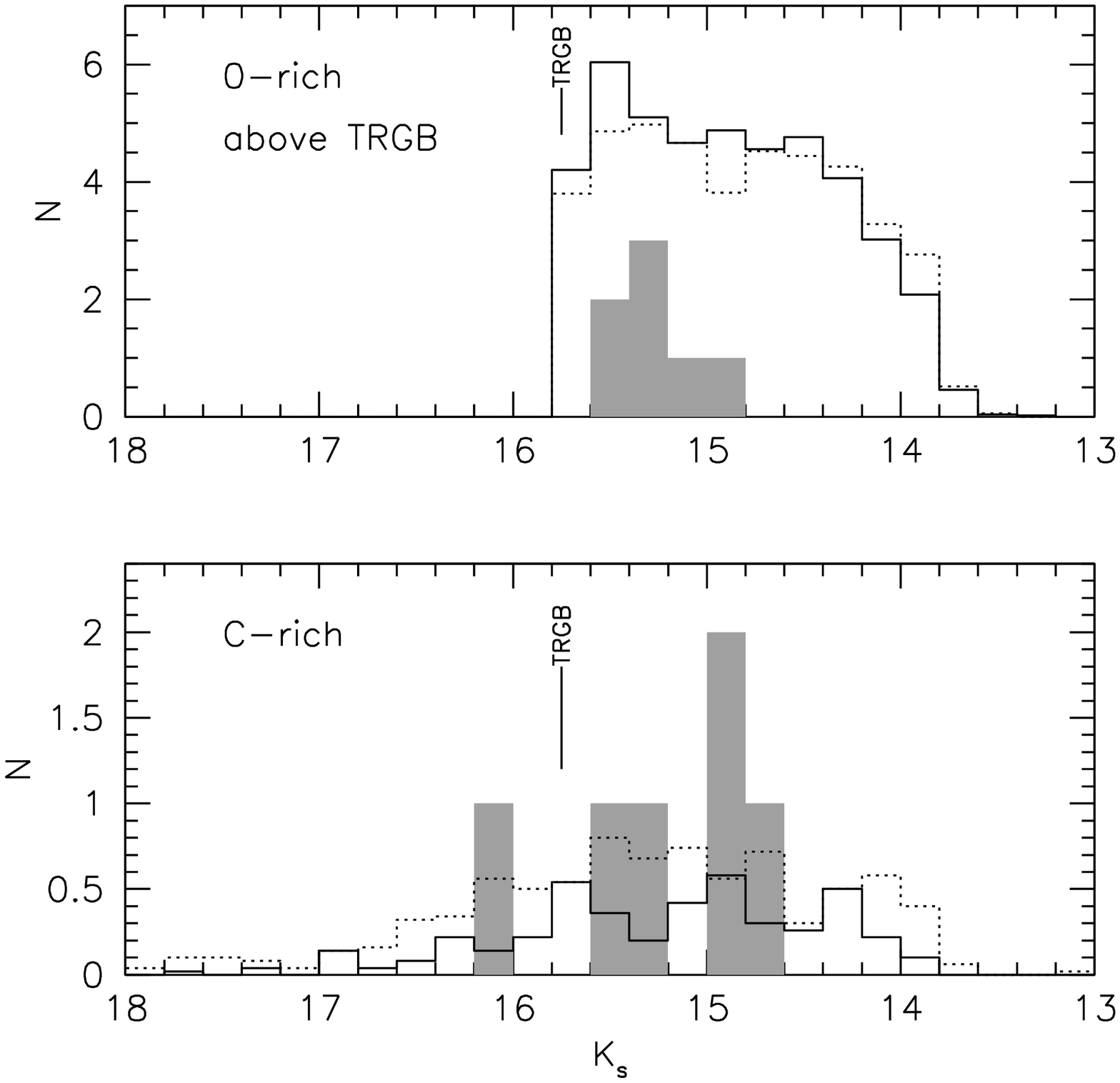}{Simulated LFs for luminous stars in
  \leoii, separated as O-rich giants above $K_{\rm s}=15.75$ 
({\it top panel}) and C-rich TP-AGB stars ({\it bottom
  panel}), for the \sfhrizzi\ ({\it solid line}) and \citet[][{\it
  dotted line}]{dolp+2005} SFHs. The {\it grey histograms} correspond to
  the stars actually observed.}{fig_simlf}

A simple comparison of Fig.~\ref{f:4cmd} and \ref{fig_simcmd} reveals
that simulated C stars have about the same $K_{\rm s}$ magnitudes 
\referee{as} 
the observed ones. This is confirmed by the bottom panel of
Fig~\ref{fig_simlf} which shows the mean C star luminosity function (as
derived from a total of 50 simulations considering the SFR errors) as
compared to the data. A KS test indicates a 59~\% probability that the
observed distribution is drawn by the predicted one, in the case of
\sfhrizzi\  SFR, and a 56~\% probability in the case
of \citet{dolp+2005}. Also, we note that the presence of a fraction of
C stars located below the TRGB is supported by present models.

The top panel of Fig~\ref{fig_simlf} shows the comparison for O-rich
giants, limited to the $K_{\rm s}=15.75$ interval so as to avoid RGB
stars. Here, it is evident that the models predict a distribution
extended to brighter magnitudes than observed. The KS probability that
the observed distribution is drawn by the predicted one is very low,
i.e. lower than 2~\% for both Rizzi's et al. and \citet{dolp+2005}
SFRs.

The discrepancy of O-rich giants is statistically significant. The
comparison with theoretical isochrones of Fig.~\ref{fig_simcmd}
suggests that the bright stars missing in the data can be identified
either with young early-AGB stars belonging to populations younger
than $\sim3$~Gyr, or to the bright section of the TP-AGB for
populations older than 7~Gyr. We have verified that eliminating any
SFR younger than 4 Gyr from the models, the problem persists. Most of
the excess of bright TP-AGB stars is produced by the old populations.

Moreover, the discrepancy in the bright part of the LF is likely
related to the excess of O-rich giants that we find in the
simulations. In TP-AGB models without third dredge-up, the total
lifetime depends essentially on the mass loss efficiency, and
particularly on the critical region of the luminosity--temperature
plane that triggers a superwind phase where most of the stellar
envelope is lost. If the superwind phase is delayed, both the lifetime
and the luminosity excursion of the TP-AGB phase would be
overestimated. Indeed, anticipating the superwind phase in O-rich
models of low mass and metallicity may constitute an interesting
solution to the discrepancy we find. 

\section{Summary and conclusions}\label{s:leo2summ} 

We have presented near-infrared $JHK_s$ photometry of a $13\farcm6
\times 13\farcm6$ field centred on \leoii\ dSph, obtained with the new
wide field imager WFCAM mounted at the UKIRT telescope.  Our data cover
most of the extension of \leoii\ dSph, and are complemented by optical
data obtained with the EMMI camera at the ESO NTT telescope.

The good statistics of our database, together with the wide colour
baseline, allowed a precise determination of the distance and
metallicity of \leoii.
We derived a distance modulus $(m-M)_0=21.68\pm0.11$ from the $J$, $H$,
and $K_s$ band magnitudes of the TRGB. This is in agreement with
optical results, confirming the reliability of our NIR methods.

The $V-K_s$ colours of RGB stars were used to derive the metallicity
distribution of the stellar populations of \leoii.  Using RGB fiducial
lines of GCs as templates, we measured a mean metallicity \mh\ $=-1.74$.
Since the bulk of the stellar population of \leoii\ is relatively old,
we estimated that the population correction to be applied to this value
is modest. Assuming a mean age of 9 Gyr yielded a correction of 0.10
dex, from which the age-corrected metallicity is \mh\ $=-1.64$.
Our measurement is in excellent agreement with recent spectroscopic
results \citep{bosl+2007,koch+2006leo2}.  A direct comparison between
spectroscopic and photometric metallicities of individual stars suggests
that the ages derived by \citet{koch+2006leo2} may be underestimated.
Indeed, older mean stellar ages would be in better agreement with the
SFHs obtained from HST photometry \citep{hern+2000,dolp2002}.

We also used our NIR data to define the properties of a nearly complete
sample of AGB stars in \leoii\ dSph.  By selecting AGB stars in the NIR
two-colour diagram, we were able to discriminate the C-rich from O-rich
stellar populations.  Foreground Milky Way stars are also easily
separated from \leoii\ AGB members.
Our NIR photometry was cross-identified with previous studies of AGB
stars, in particular C stars \citep{azzo+1985}.  One of the 7 carbon
star listed by \citet{azzo+1985} has anomalous colours, and visual
inspection of our images confirms that it is a background galaxy. No
indication for additional C stars above the TRGB was found in our
analysis, therefore we conclude that the remaining 6 stars represent the
complete population of C stars in \leoii\ within the area covered by our
observations.  Using our colour selection, we provide the first sample
of O-rich stars in \leoii\ above the TRGB, with negligible contamination
from foreground Milky Way stars.

Our \leoii\ observations were modeled via simulations based on the
HST-derived SFHs, and using the most updated isochrones.
The comparison between data and simulations has evidenced both
successful and discrepant points, which are all potentially important
for the calibration of AGB star models at low metallicity. 

With respect to the O-rich TP-AGB stars, the most important discrepancy
consists in a predicted over-estimation of their number and mean
$K_s$-band luminosities, as compared to the data.  Interestingly,
\citet{will+2007} find similar indications of an excess of AGB stars in
the oldest \citet{gira+2000} isochrones while fitting RGB and AGB stars
in the Virgo intracluster stars 
observed with HST/ACS. A possible solution to
this problem could be an increase of mass loss efficiency in O-rich
TP-AGB models of low metallicity and low mass. This will be explored in
forthcoming work. 
Regarding this point, we note that:
\begin{enumerate}
\item The obscuration of AGB stars by circumstellar dust could be much
more efficient than assumed here for low metallicity, and contribute
to the solution of this problem. Indeed, recent Spitzer/IRAC
observations of the dwarf irregular galaxies WLM and IC\,1613
\citep{jack+2007a,jack+2007b} indicate a very high fraction of
optically-obscured AGB stars, of about 40~\%. In order to reduce the
discrepancies in the LFs for \leoii, dust obscuration should be
affecting mainly the O-rich AGB stars of higher luminosities.

\item This discrepancy could still be reduced by adopting alternative
SFHs. It would be again desirable to have improved derivations of the
SFH in dwarf galaxies, based on HST observations covering larger areas
than those used by \citet{dolp+2005} and \sfhrizzi.
Progress in this sense is expected as the result
of ongoing HST surveys and legacy programs on dwarf galaxies
\citep{dolp+2005,dalc2006,gall+2007}.
\end{enumerate}
The above-mentioned discrepancy is still based on too a small sample of
observed AGB stars. In forthcoming papers we will extend the comparison
to other dwarf galaxies in the Local Group, in order to increase the
statistical significance of our conclusions.

With respect to the C-rich TP-AGB stars, the comparison between data and
simulations is satisfactory, in terms of location in the
colour-magnitude diagram, counts, and luminosity functions. We derive
the following important indications:
\begin{enumerate}
\item At low metallicity the minimum mass for a star to become a C~star 
can be as low as $\sim\,1\, M_\odot$, which sets an important
constraint to the treatment of the third dredge-up in AGB stellar
models.
\item Low-mass C stars produced by the TP-AGB evolution of single
stars (i.e. not belonging to binary systems), are expected to populate
the O-rich sequence of giants below the TRGB when they 
are in the long-lived low-luminosity dips driven by thermal
pulses \citet{boot+sack1988}.
\item The higher effective temperatures of C~stars with bluer \jks\
colours can be a consequence of (a) less efficient molecular formation
and opacity at lower metallicity, and/or (b) warming of their Hayashi
line during the low-luminosity stages of pulse cycles
\citep[see][]{marigira2007}.
\end{enumerate}

From our analysis, we conclude that constraints to the TP-AGB models can
be obtained from dwarf galaxies with known SFHs, provided that the
numbers of AGB stars are significant and that the SFH is known with
sufficient accuracy.

\section*{Acknowledgments}

We warmly thank M. Riello for helpful comments and support with the WFCAM
pipeline, M.A.T. Groenewegen for his help in setting the {\sc trilegal} code,
and A. Dolphin for providing SFH data ahead of publication.  We
acknowledge support to this project by the Italian MUR through the PRIN
2002028935 (P.I. M. Tosi) and PRIN 2003029437 (P.I. R. Gratton)
Projects, and by the University of Padova (Progetto di Ricerca di Ateneo
CPDA052212).
The United Kingdom Infrared Telescope is operated by the Joint Astronomy
Centre on behalf of the Science and Technology Facilities Council of the
U.K.
This publication made use of data products from the Two Micron All Sky
Survey, which is a joint project of the University of Massachusetts and
the Infrared Processing and Analysis Center/California Institute of
Technology, funded by the National Aeronautics and Space Administration
and the National Science Foundation.

\label{lastpage}

\end{document}